\documentclass[twocolumn,useAMS,usenatbib]{mn2e}
\usepackage{graphicx}
\usepackage{amsmath}
\usepackage{amssymb}
\usepackage{lscape}
\usepackage{color}
\usepackage{bm}
\usepackage{dsfont}
\usepackage{afterpage}
\usepackage{float}

\def\be{\begin{equation}}
\def\ee{\end{equation}}
\def\bea{\begin{eqnarray}}
\def\eea{\end{eqnarray}}

\def\wde{w_{de}}

\newcommand{\h}{\mathcal{H}}
\newcommand{\comment}[1]{}

\title{Observational constraints on an interacting dark energy model}

\author[Jussi V\"{a}liviita, Roy Maartens and Elisabetta Majerotto]{Jussi V\"{a}liviita,$^{1}$ Roy Maartens$^{1}$ and Elisabetta Majerotto$^{2,1}$\\
$^{1}$Institute of Cosmology \& Gravitation, University of Portsmouth, Portsmouth PO1 3FX, United Kingdom\\
$^{2}$INAF-Osservatorio Astronomico di Brera, Via Bianchi 46, I-23807 Merate (LC), Italy}

\begin{document}
\label{firstpage}

\date{Accepted 2009 November 25. Received 2009 November 17; in original form 2009 September 19}
%\date{Received in original form 2009 September 18}
\pagerange{\pageref{firstpage}--\pageref{lastpage}} \pubyear{20XX}

\maketitle

\begin{abstract}

We use observations of cosmic microwave background
anisotropies, supernova luminosities
and the baryon acoustic oscillation signal in the galaxy distribution to constrain the
cosmological parameters in a simple interacting dark energy model
with a time-varying equation of state.
Using a Monte Carlo Markov Chain technique we determine the posterior
likelihoods. Constraints from the individual data sets are weak, but the
combination of the three data sets confines the interaction
constant $\Gamma$ to be less than 23\% of the expansion
rate of the Universe $H_0$; at 95\% CL $-0.23 < \Gamma/H_0 < +0.15$.
The CMB acoustic peaks can be well fitted even if the interaction rate
is much larger, but this requires a larger or smaller (depending on the
sign of interaction) matter density today than in the non-interacting model.
Due to this degeneracy between the matter density and the interaction rate,
the only observable effect on the CMB is a larger
or smaller integrated Sachs--Wolfe effect.
While SN or BAO data alone do not set any direct constraints
on the interaction, they exclude the models with very large
matter density, and hence indirectly constrain the interaction
rate when jointly analysed with the CMB
data. To enable the analysis described in this paper,
we present in a companion paper [arXiv:0907.4981]
a new systematic analysis of the early
radiation era solution to find the adiabatic initial
conditions for the Boltzmann integration.

\end{abstract}

%\pacs{98.70.Vc, 98.80.Cq}


\section{Interacting dark energy}

Dark energy and dark matter are the dominant sources in the
`standard' model for the evolution of the universe. Both are
currently only detected via their gravitational effects, with an
inevitable degeneracy between them (one requires a model to
separate dark energy from dark matter). There could thus be an
interaction between them that is consistent with current observational
constraints. A dark sector interaction could also alleviate the
`coincidence' problem (why are the energy densities of the two
components of the same order of magnitude today?). Furthermore,
interacting dark energy exerts a non-gravitational `drag' on dark
matter, and thus can introduce new features to structure
formation, including possibly a new large-scale
bias~\citep{Amendola:2001rc} and a violation by dark matter of the
weak equivalence principle on cosmological
scales~\citep{Bertolami:2007zm,Koyama:2009gd}.

The energy balance equations in the background are
\bea
\label{eq.cont_rhoc} \rho_c' &=& -3 \h \rho_c + a Q_c\,, \\
\rho_{de}' &=& -3 \h (1+\wde) \rho_{de} + a Q_{de}\,,\quad
Q_{de}=-Q_c\,, \label{eq.cont_rhode}
\eea
where $a$ is the scale factor of the Universe, $\h = a'/a$  is the conformal Hubble parameter,
$\wde=p_{de}/\rho_{de}$ is the dark energy equation of state parameter,
a prime indicates derivative with respect
to conformal time $\tau$, and $Q_c$ is the rate of transfer to the
dark matter density due to the interaction.

Various forms for $Q_c$ have been put forward (see,
e.g.~\citet{Wetterich:1994bg,Amendola:1999qq,Billyard:2000bh,Zimdahl:2001ar,Farrar:2003uw,Chimento:2003iea,Olivares:2005tb,Koivisto:2005nr,Sadjadi:2006qp,Guo:2007zk,Boehmer:2008av,He:2008tn,Quartin:2008px,Pereira:2008at,Quercellini:2008vh,Valiviita:2008iv,He:2008si,Bean:2008ac,Chongchitnan:2008ry,Corasaniti:2008kx,CalderaCabral:2008bx,Gavela:2009cy,Jackson:2009mz}).
All of these models are phenomenological. Some of them are
constructed specifically for mathematical simplicity -- for
example, models in which $Q_c \propto {\cal H}\rho$. Rather than
design the interaction to achieve a specific outcome, we prefer to
start with a simple physical model, and then develop its
predictions.

\begin{figure*}
\centering
\includegraphics[width=0.45\textwidth]{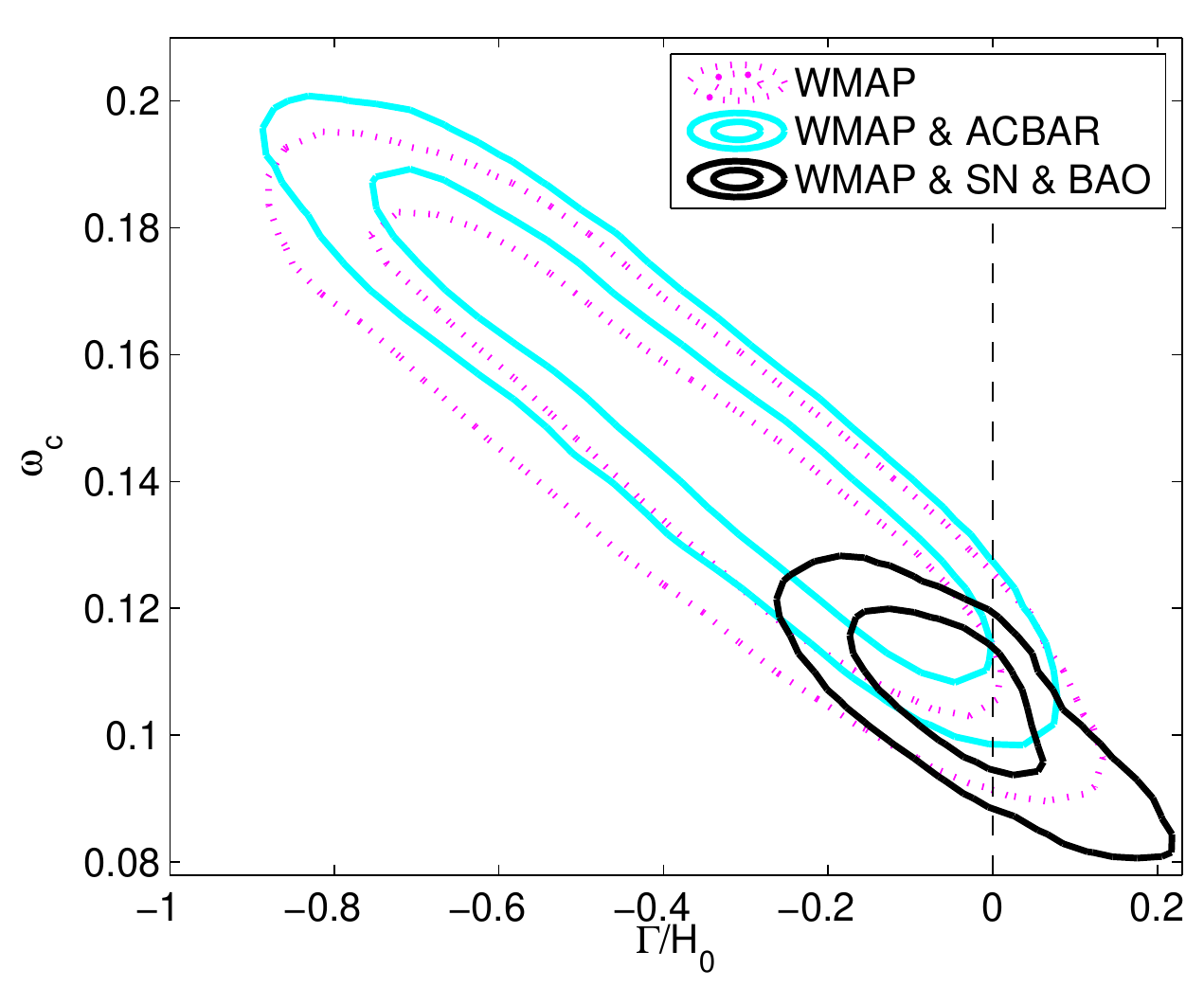}
\caption{ Marginalized likelihoods for the interacting model with
WMAP, WMAP\&ACBAR, and WMAP\&SN\&BAO data. The curves show 68\%
and 95\% CL regions: dotted magenta/grey for WMAP, solid cyan/grey
for  WMAP\&ACBAR, and solid black for  WMAP\&SN\&BAO data.
\label{fig:Like2docGamma}}
\end{figure*}

We consider models which are similar to simple models of
reheating~\citep{Turner:1983he}, of dark matter decay into
radiation~\citep{Cen:2000xv}, and of curvaton
decay~\citep{Malik:2002jb} -- i.e., where the interaction has the
form of a decay of one species into another, with constant decay
rate. Such a model was introduced
in~\citet{Boehmer:2008av,Valiviita:2008iv}: \be Q_c =- \Gamma
\rho_c \,,\label{eq:ourcoupling} \ee  where $\Gamma$ is the
constant positive decay rate of dark matter to dark energy. Here,
as well as in \citet{Valiviita:2008iv},
we include also the possibility $\Gamma<0$, i.e.,
allow also for energy transfer is from dark energy to dark matter.

In \citet*{Valiviita:2008iv} we considered the case of fluid dark
energy with a \emph{constant} equation of state parameter $-1 < \wde \le
-4/5$, and found that a serious large-scale non-adiabatic
instability affects this model in the early radiation dominated
epoch. This instability is stronger the closer $\wde$ is to $-1$.
Phantom models, $\wde < -1$, do not suffer from this instability,
but we consider them to be unphysical.

The instability is determined by the early-time value of
$\wde$: for constant $\wde$ there is no flexibility, but a
\emph{variable} $\wde$ (as in quintessence models, see e.g. \citet{Lee:2006za})
should allow us to avoid the
instability. We show that the models are viable if $w_{de}> -4/5$ at
early times, while at late times, $\wde \sim -1$.
We represent $\wde$ via the parametrization $ \wde = w_0 + w_a
(1-a)$ \citep{Chevallier:2000qy,Linder:2002et}, which we rewrite
as \be \wde = w_0a + w_e (1-a)\,, \label{eq:wdeedef} \ee where
$w_e =w_0+w_a$ is the early-time value of $\wde$, while $w_0$ is
the late-time value. This parametrization was originally developed
to phenomenologically account for the possible time evolution
of $\wde$ up to redshifts of a few. At higher redshifts it describes
the simplest possible model where $\wde$ is a constant, namely $w_e$.
In particular, this parametrization works  well for some classes of quintessence models \cite{Caldwell:2005tm} and it has the advantage of having a finite value at high redshifts. Using a parametrisation for $\wde$ has the drawback that it cannot reproduce all possible models for a large redshift range. Furthermore, it is not possible to compute the speed of sound but it is necessary to assume a value for it. It might be interesting to look at different parametrisations of $\wde$ or a definite scalar field model. This is left for future work. In this paper we demonstrate with the simple parametrization, Eq.~(\ref{eq:wdeedef}), that a (suitably) time-varying $\wde$ cures the interacting model, Eq.~(\ref{eq:ourcoupling}), from the early-time large-scale instability, and thus provides a viable cosmology.

We perform a full Monte Carlo Markov Chain (MCMC) likelihood scan
for the spatially flat interacting and non-interacting models,
using a modification of the CAMB
%\footnote{http://camb.info/}
\citep{Lewis:1999bs}
Boltzmann code, to determine the
best-fitting values of $\Gamma/H_0$ and the other cosmological
parameters, against
Wilkinson Microwave Anisotropy Probe (WMAP) 5-years data \citep{Komatsu:2008hk},
Supernovae Ia (SN) Union sample data \citep{Kowalski:2008ez}, and
data points for the ratio of the sound horizon to a distance measure
at two different redshifts from  baryon acoustic oscillation
(BAO) observations \citep{Percival:2007yw}.
The best-fitting models with various combinations of data are shown in Table~\ref{table:Models}
on page \pageref{app:table} in Appendix~\ref{app:table}, while Fig.~\ref{fig:Like2docGamma} above
summarizes our main findings for the posterior likelihoods.
Our most stringent results for the interacting model
result from the combined analysis of WMAP\&SN\&BAO, giving
the following minimal 95\% intervals:
% \bea
%&& \Gamma/H_0 \in (-0.23,\, +0.15), ~~ w_e \in (-0.80,\, -0.19),
%~~ w_0 \in (-1.00,\, -0.63), \\ && \omega_b \in (0.0212,\,
%0.0241), ~~\omega_c \in (0.859, \, 0.125),~~ \Omega_{de0} \in
%(0.648,\, 0.767),~~ H_0 \in (63,\, 70),\\
%&& n_S \in (0.937,\, 1.002), ~~ \ln(10^{10}A_S^2) \in (2.95,\,
%3.14), ~~\tau_{re} \in (0.057,\, 0.133), ~~ \mbox{Age }\in
%(13.6,\, 14.3)\,\mbox{Gyr}.
% \eea
$\omega_b \in (0.0212,\,
0.0241)$, $\omega_c \in (0.859, \,  0.125)$, $H_0 \in (63,\, 70)$,
$\tau \in (0.057,\, 0.133)$, $\Gamma/H_0 \in (-0.23,\, +0.15)$,
$w_e \in (-0.80,\, -0.19)$, $w_0 \in (-1.00,\, -0.63)$, $n_S \in
(0.937,\, 1.002)$, $\ln(10^{10}A_S^2) \in (2.95,\, 3.14)$,
$\Omega_{de0} \in (0.648,\,  0.767)$, Age$\,\in (13.6,\,
14.3)\,$Gyr.
Description and prior ranges of these parameters are given
in Appendix~\ref{app:ranges}. There, in Table A1,
we also list the definitions of some of the symbols used in this paper.

\begin{table*}
\caption{The evolution of perturbations on super-Hubble scales with various values of the dark energy equation of state parameter in the radiation and matter dominated eras. ``Adiabatic'' means that it is possible to specify adiabatic initial conditions so that the total gauge invariant curvature perturbation $\zeta$ stays constant on super-Hubble scales, and the evolution of all non-dark energy perturbations is the same as in the non-interacting case while the $de$ perturbations behave differently. ``Adiabatic (standard)'' means that the behaviour of all perturbations at early times on super-Hubble scales is the same as in the non-interacting model.\label{table:summary}}
  \begin{tabular}{|l|l|l|l|}
  \hline
  $w_{de}$ in the RD or MD era & Radiation dominated era (RD) & Matter dominated era (MD) & Viable? \\
 \hline
   $w_{de}<-1$ & adiabatic & adiabatic & viable, but phantom \\
   $-1< w_{de} < -4/5 $ & ``blow-up'' isocurvature growth & ``blow-up'' isocurvature growth & non-viable \\   $ -4/5 \le w_{de} < -2/3$ & adiabatic &  isocurvature growth & viable, if $|\Gamma|$ small enough \\
   $ -2/3 \le w_{de} < -1/2$ &  adiabatic (standard) &  adiabatic & viable \\
   $ -1/2 \le w_{de} < +1/3 $ & adiabatic (standard) &  adiabatic (standard) & viable \\
\hline
  \end{tabular}
\end{table*}

The key features of the constraints on the interacting model from
data may be summarized as follows.
\begin{itemize}
\item
Any of the data sets alone (CMB, or SN, or BAO) would allow for
large interaction: $|\Gamma|$ could be even larger than
today's Hubble rate $H_0$. In the CMB the only hint
from a large interaction rate is a modified integrated Sachs--Wolfe (ISW)
effect. However, due to the cosmic variance the $\chi^2$ is only mildly
affected.
%The acoustic peaks in the CMB are not sensitive to the interaction
%-- only the ISW is affected. The SN and BAO are sensitive to the
%interaction via the change in $\omega_c$ (or equivalently
%$\Omega_{de0}$).
\item
A large negative $\Gamma$ fits the CMB TT and TE spectra equally
well as the $\Gamma=0$ model, but a good fit requires a larger
physical cold dark matter density today, $\omega_c$,
(and hence a smaller $\Omega_{de0}$) and a smaller
$H_0$. Negative $\Gamma$ suppresses the late ISW effect and hence the
CMB fit is slightly better than in the $\Gamma=0$ case. Combining
the CMB data with either SN or BAO or SN\&BAO data, this
improvement is cancelled by a worse fit to SN and BAO due to too
little acceleration at low redshifts.
%The opposite holds for a positive $\Gamma$.
\item
Models with large positive $\Gamma$ fit the high-$l$ CMB TT and
all TE data equally well as the $\Gamma=0$ model, but a good fit
requires a smaller $\omega_c$ (and hence a larger $\Omega_{de0}$)
and a larger $H_0$. The SN and BAO data can be fitted better than
in the $\Gamma=0$ case due to increased acceleration at small
redshifts. However, this improvement is cancelled by a worse fit
to the low-$l$ CMB TT spectrum due to a large late ISW effect.
\end{itemize}

There are two critical features of the analysis of interacting
models, which are not always properly accounted for in the
literature:
\begin{itemize}
\item
The background energy transfer rate $Q_c$ does not in itself
determine the interaction in the perturbed universe: one must also
specify the momentum transfer rate. We do this via a physical
assumption, i.e., that the momentum transfer vanishes in the dark
matter rest-frame, so that the energy-momentum transfer rate is given
covariantly \citep{Valiviita:2008iv} by
 \be \label{qcmu}
Q_c^\mu = Q_c u_c^\mu = -Q_{de}^\mu\,,\quad Q_c=-\Gamma \rho_c(1+
\delta_c),
 \ee
where $u_c^\mu$ is the dark matter 4-velocity, and $\delta_c=\delta\rho_c/\rho_c$ is the
cold dark matter (CDM) density contrast.
\item
Adiabatic initial conditions in the presence of a dark sector interaction
require a careful analysis of the early-radiation solution. We
derive these initial conditions in the companion paper \citep*{CompanionPert}
by generalizing the methods of \citet{Doran:2003xq} to the interacting case, extending
our previous results \citep{Valiviita:2008iv}. The key
results for the initial conditions and early-time
perturbation evolution are reproduced in Table~\ref{table:summary}.
\end{itemize}

We give here the first analysis of the CMB spectra and the first
MCMC likelihood analysis for the interacting model~(\ref{qcmu}),
using the perturbation equations and initial conditions
given in the companion paper \citep*{CompanionPert}.
Cosmological perturbations of other interacting models have been
investigated in
\citet{Amendola:2002bs,Koivisto:2005nr,Olivares:2006jr,Mainini:2007ft,Bean:2007ny,Vergani:2008jv,Pettorino:2008ez,Schaefer:2008qs,Schaefer:2008ku,LaVacca:2008kq,He:2008si,Bean:2008ac,Corasaniti:2008kx,Chongchitnan:2008ry,Jackson:2009mz,Gavela:2009cy,LaVacca:2009yp,He:2009mz,CalderaCabral:2009ja,He:2009pd,Koyama:2009gd,Kristiansen:2009yx}.

%\section{MCMC likelihood scan of the interacting model}

%The crucial ingredient of the analysis, the initial conditions deep in the radiation era,
%are derived in the companion paper \citep{x} where we show that

\section{Phenomenology}
\label{sec:phenomenology}

We have performed 7 MCMC runs for the spatially flat ($\Omega=1$)
interacting model with various data sets (WMAP, WMAP\&ACBAR, SN,
BAO, WMAP\&SN, WMAP\&BAO, and WMAP\&SN\&BAO). Here WMAP refers to
the 5-year temperature and polarization anisotropy data \citep{Komatsu:2008hk},
ACBAR to the Arcminute Cosmology Bolometer Array Receiver  data \citep{Reichardt:2008ay},
SN to the Union Supernovae Ia sample \citep{Kowalski:2008ez} as implemented in
CosmoMC\footnote{http://cosmologist.info/cosmomc}  \citep{Lewis:2002ah,cosmomc_notes}
with systematic errors flag turned on, and BAO to the two
data points $r_s(z_{\rm dec})/D_V(z=0.2)$ and $r_s(z_{\rm
dec})/D_V(z=0.35)$ from \citet{Percival:2007yw}.
For reference we have done also 6
MCMC runs (excluding WMAP\&ACBAR from the above list) for the
spatially flat non-interacting model. Each of these 13 runs has
3--4 chains with mean input multiplicity in the range 3--10, and
the number of accepted models in each chain is $\sim$25000. The
measure of mixing, the worst eigenvalue $R-1$ (which is better the
closer it is to zero), is for all cases less than $0.03$.
More technical details are given in Appendix~\ref{sec:technicaldetails}.
%In this section we discuss the phenomenology of the model,
%give the posterior likelihoods, discuss which additional data could
%set more stringent constraints on the model, and finally conclude.

\begin{figure*}
\centering
\includegraphics[width=\textwidth]{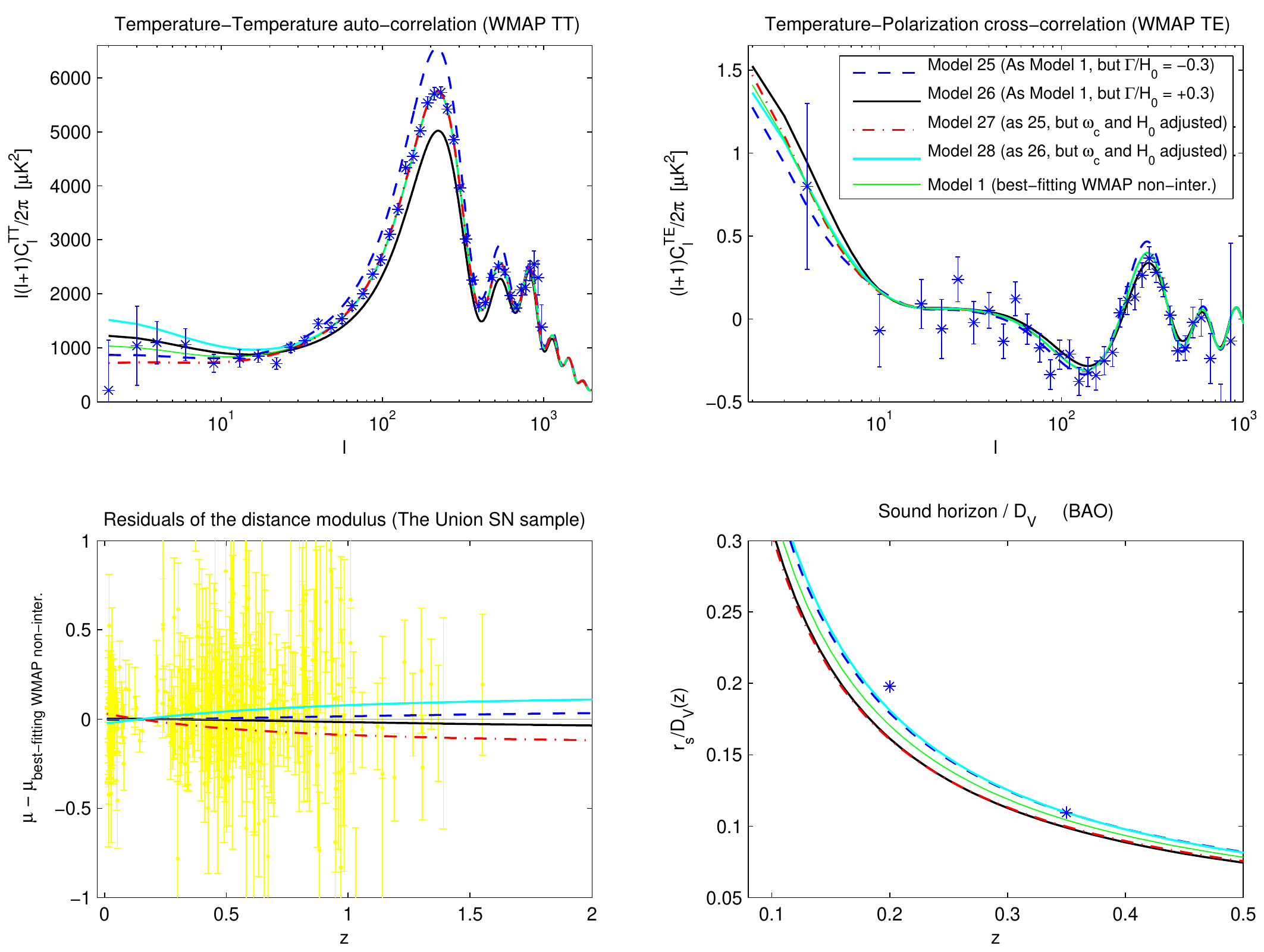}
\caption{Models with most of their parameters equal to the
parameters of the best-fitting to WMAP non-interacting model.
%Note: In all figures 'coupled' refers to the interacting model ($\Gamma\neq 0$), 'uncoupled' to the non-interacting model ($\Gamma=0$), and $w_{de}^e$ to the early-time dark energy equation of state parameter which we call $w_e$ in the text.
\label{fig:OtherModels}}
\end{figure*}

Table~\ref{table:Models}  on page \pageref{app:table}
in Appendix~\ref{app:table} collects the
best-fitting models of each run
(excluding WMAP\&ACBAR). In addition, we show for each data-set
two relatively good-fit models with a strong negative/positive
interaction ($|\Gamma/H_0| > 0.1$) which are not 'far'
from the best-fitting non-interacting model -- in a sense
that $\Delta\chi^2 = \chi^2$(strong interaction model) --
$\chi^2$(best-fitting model) $<4$. In the last four rows of the table
(Models 25--28) we show interacting models which have most of
their parameters equal to the WMAP best-fitting non-interacting model
(Model 1).

Fig.~\ref{fig:OtherModels} shows the angular power spectra,
distance modulus, and $r_s(z_{\rm dec})/D_V(z)$ for Models 25--28
and Model 1 from Table~\ref{table:Models}. The WMAP best-fitting
non-interacting model (Model 1; thin green/grey line) fits well all the
data. The only exceptions are that it fails to fit the low
quadrupole $l=2$ in the WMAP TT spectrum, and undershoots both
the $z=0.20$ and $z=0.35$ BAO data points, whose error bars are
smaller than the asterisk symbol in the plot. Therefore it would
be surprising if the interacting model could fit the data
overall much better. Now we take Model 1 and turn the interaction
on to $|\Gamma/H_0| = 0.3$ (Models 25 and 26). The negative
(positive) interaction leads to an extremely bad fit to the CMB TT
data, as the the model now vastly overshoots (undershoots) the
first and second acoustic peaks. This is also reflected in the TE
spectrum where the peak at $l\sim300$ is overshot
(undershot). The SN data cannot distinguish the interacting model
from the non-interacting model, though there is a tiny difference:
a negative (positive) interaction leads to a faster (slower)
increase of the distance modulus as a function of redshift than
seen in the non-interacting model, and hence to a slightly better
(worse) fit to the SN data. Finally, a negative interaction helps
fitting the BAO data as it systematically shifts $r_s(z_{\rm
dec})/D_V(z)$ upward. A positive interaction makes the fit to BAO
data worse. All these remarks are reflected in the
$\chi^2$ values presented in Table~\ref{table:Models}.

It is quite straightforward to understand why a negative
interaction helps in fitting the SN and BAO data. The negative
interaction means that the energy transfer is from dark energy to
dark matter. As we have kept today's values of $\omega_c$ (and
$\Omega_{de0}$) and $H_0$ fixed, this means that in the negatively
interacting model there has been more dark energy in the past than
in the non-interacting model. This causes more acceleration of the
expansion of the universe, and hence larger distance moduli at
high redshifts. In the BAO the distance measure is in the
denominator, and at first sight one would expect smaller
$r_s(z_{\rm dec})/D_V(z)$ than in the non-interacting case.
However, as $D_V$ is proportional to $D_A^{1/3}$ the effect of
an increased angular diameter distance, $D_A(z)$, is mild.
Instead the main effect now comes from
the different sound horizon. For a fixed today's $\omega_c$ the
negatively interacting model has a smaller cold dark matter
density at last scattering $\omega_c(z_\ast)$ than the
non-interacting model. The sound horizon $r_s(z_{\rm dec})$ is
proportional to $k_{\rm eq}^{-1}$ where $k_{\rm eq}$ is the wave
number corresponding to the scale $\lambda_{\rm eq}$ that
re-enters the horizon at matter-radiation equality. The smaller
$\omega_c(z)$ at early times means that the matter density exceeds
the radiation energy density later than in the non-interacting
case. As small scales re-enter the Hubble horizon before the large
scales, $\lambda_{\rm eq}$ (or $k_{\rm eq}^{-1}$) and hence the
sound horizon $r_s(z_{\rm dec})$ will be larger in the negatively
interacting case. Table~\ref{table:Models} confirms these
conclusions (compare $D_V$ and $r_s$ in Models 1 and 25). The
positive interaction model behaves vice versa.

Now we can explain also the bad fit to the CMB. The angular power
spectrum is very sensitive to the dark matter density at
last scattering $\omega_c(z_\ast)$ (or at the redshift of
matter-radiation equality $z_{\rm eq}$). As the negative interaction
model has a small $\omega_c(z_\ast)$ and hence a small
$z_{\rm eq}$, the matter-radiation equality occurs very close to last
scattering $z_\ast$. Therefore last scattering does not happen
in the matter dominated era but around the transition from
radiation domination to matter domination. This causes a large
early integrated Sachs--Wolfe effect (eISW), which amplifies the
first (and second) acoustic peaks. The angular power spectrum
looks like that of a non-interacting model with a very small
today's $\omega_c$. The positive interaction model looks like a
non-interacting model with a very high  today's $\omega_c$, i.e.,
the first and second acoustic peaks are vastly undershot (due to
the early matter-radiation equality and hence a small eISW
amplification). Therefore, in order to obtain a good fit to the CMB in
the interacting model, we need to adjust today's $\omega_c$ in such
a way that $\omega_c(z_\ast)$, or rather $a_{\rm eq}$, matches with the
best-fitting non-interacting model. We have done this in Models 27 and
28; see Fig.~\ref{fig:OtherModels} and Table~\ref{table:Models}.
In the best-fitting non-interacting model (Model 1) today's dark
matter density is $\omega_c = 0.107$, whereas to obtain a good fit
with a negative (positive) interaction of $\Gamma/H_0 = -0.3$
$(+0.3)$ we require a larger $\omega_c=0.137$ (smaller
$\omega_c=0.080$). As the larger (smaller) dark matter density
leads to a smaller (larger) $\Omega_{de0}$, the interacting model
will now have a smaller (larger) angular diameter distance to last
scattering $D_A(z_\ast)$. This would lead to all the acoustic peak
structure shifting slightly to the left (right) from the WMAP data.
As $D_A$ is proportional to $H_{0}^{-1}$, this mismatch can be
corrected by decreasing (increasing) the value of $H_0$ so that we
obtain roughly the same $D_A(z_\ast)$ and hence the same
acoustic peak positions as in the non-interacting case. With  a
negative (positive) interaction of $\Gamma/H_0 = -0.3$ $(+0.3)$, we
require $H_0 = 56$ ($66$), while the best-fitting non-interacting model
has $H_0=61\,$km$\,$s$^{-1}$Mpc$^{-1}$. (Note: $D_V$ and $r_s$ in
Table~\ref{table:Models} are reported in units $h^{-1}$Mpc. The
interacting Models 27 and 28 have $r_s = 154\,$Mpc, which is
exactly the same as for the non-interacting Model 1.)

Fig.~\ref{fig:OtherModels} and Table~\ref{table:Models} show that
after the above-described adjustments the strongly interacting
$|\Gamma/H_0|=0.3$ Models 27 and 28 provide an excellent fit to
the CMB acoustic peaks. In the high-$l$ ($l>32$) region of the TT
and all of the TE spectrum, the interacting models are totally
indistinguishable from the non-interacting best-fitting Model 1.
However, as the negatively interacting Model 27 has small
$\Omega_{de0} = 0.49$ and hence also a smaller $\Omega_{de}$ in
the recent past than the non-interacting model, the late
integrated Sachs--Wolfe effect is suppressed,
so there is less TT power at low multipoles. As the
non-interacting model slightly over shoots the $C_l^{\rm TT}$
spectrum at low multipoles, the negatively interacting model leads
to a better fit here; see Fig.~\ref{fig:OtherModels} upper left
panel and the column $\chi^{2\rm TT}_{l<32}$ in
Table~\ref{table:Models} for Models 1 and 27. (Note: the
$\chi^{2\rm TT}_{l<32}$ numbers are from Gibbs sampling of the
actual CMB map, and the more negative the number is the better the
fit is.) The positive interaction Model 28 has a larger
$\Omega_{de}$ in the recent times
than the non-interacting model, and hence
gives rise to a large late ISW effect, and a poor fit to low
multipoles in the TT spectrum.

% moved old fig 3 from here

The SN data cover the relatively recent past only and hence the small
$\Omega_{de0}=0.49$ in the negative interaction Model 27
leads to a poorer fit. The SN data indeed slightly favour more
acceleration in the recent past and the positive interaction Model
28 ($\Omega_{de0}=0.76$) has this property, leading to a better fit
to SN than the non-interacting Model 1, which has
$\Omega_{de0}=0.65$; see Fig.~\ref{fig:OtherModels} lower left
panel and the column $\chi^{2}_{\rm SN}$ in
Table~\ref{table:Models}. For the same reason the negative
interaction model fits the BAO data worse than the positive
interaction model. Also the BAO data cover only the relatively
recent past and favour more acceleration between redshifts
$z=0.20$ and $z=0.35$ than the best-fitting non-interacting model
provides.

Indeed already from Table~\ref{table:Models} we can see these
general trends. All the best-fitting interacting models (2, 14, 18,
and 22) have negative $\Gamma$ and are slightly better fits to
the CMB due to sightly better fit to low-$l$ CMB spectra. In general
the best-fitting interacting models have indistinguishable
$\chi^2_{\rm SN}$ and $\chi^2_{\rm BAO}$ from the best-fitting non-interacting
models. The largest-$|\Gamma|$ models within $\Delta\chi^2 < 4$
from the best-fitting model (see Models 3, 4, 15, 16, 19,
20, 23, 24 in Table ~\ref{table:Models}) always obey the
following: a good-fit large negative interaction model is by about
$\Delta\chi^2=1$--$3$ better fit to WMAP than a good-fit large
positive interaction model, again due to the low-$l$ behaviour.
However, a negative interaction model is always a worse fit to
both SN and BAO data (due to too small $\Omega_{de}$ in the recent past).

%%%%%%%%%%%%%%%%%%%%%%%%%%%%%%%%%%%%%%%%%
\begin{figure*}
\centering
\includegraphics[width=\textwidth]{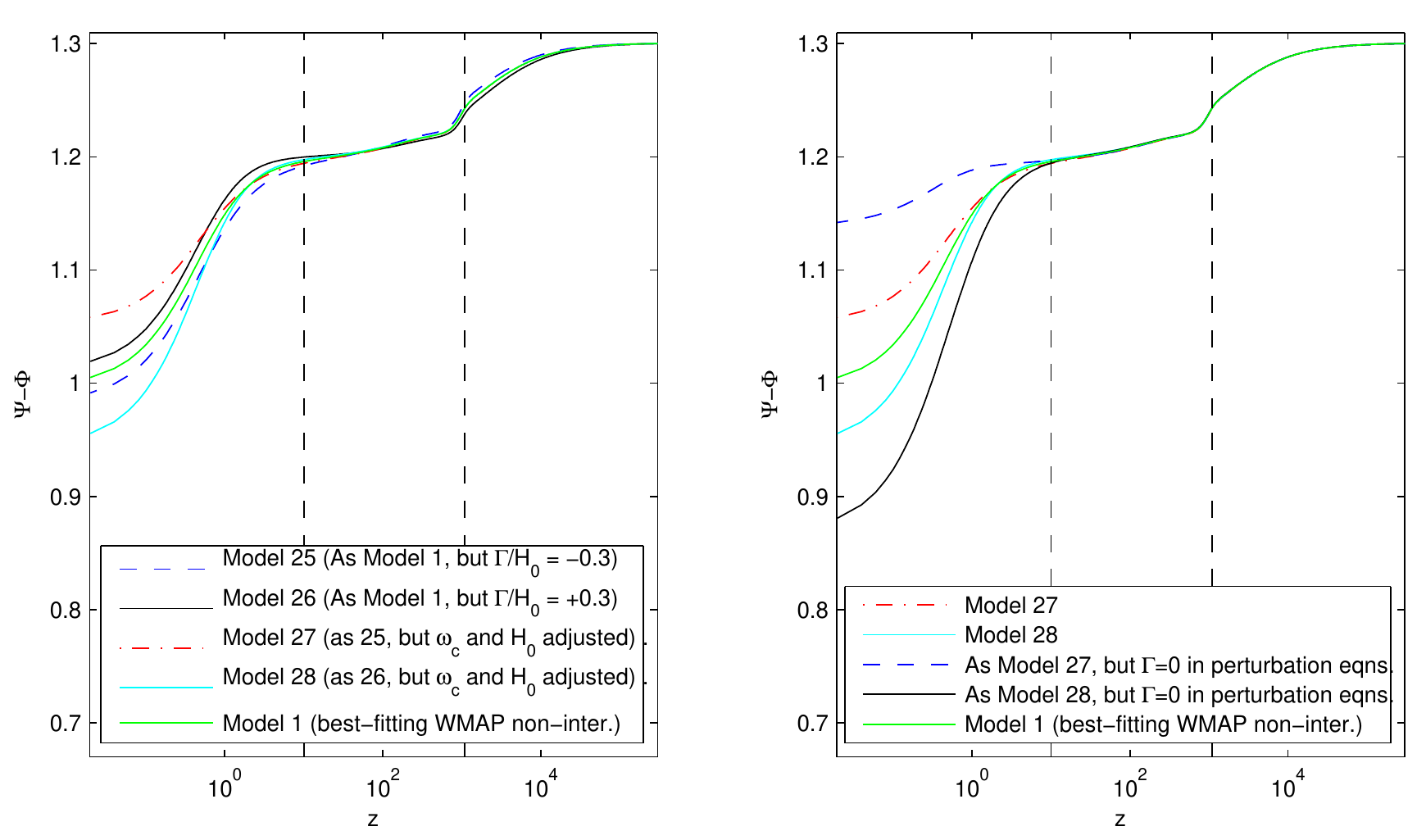}
\caption{Redshift evolution of the ISW source. Dashed
vertical lines indicate last scattering ($z \approx 1090$) and the
time when the interaction starts to directly modify the evolution
of the ISW source ($z \approx 10$). \label{fig:ISWsource}}
\end{figure*}
%%%%%%%%%%%%%%%%%%%%%%%%%%%%%%%%%%%%%%%%%

The ISW effect in the temperature angular power spectrum is given
by
\bea
%\textstyle
C_l^{ISW} & = & 4\pi \int_0^\infty \frac{k^3}{2\pi^2} \bigg\{
\int_{0}^{\tau_0} \Big[ \Big(\Psi'(\tau,k)-\Phi'(\tau,k)\Big)\nonumber\\
&& \times
j_l\Big(k(\tau_0-\tau)\Big) e^{-\tau_{\mathrm{od}}(\tau,\tau_0)}
\Big] d\tau \bigg\}^2 \frac{dk}{k}\,,
\eea
where $j_l$ is the spherical Bessel function and $\tau_{\mathrm{od}}(\tau,\tau_0)=\int_{\tau}^{\tau_0}\mbox{opacity}(\tilde\tau)d\tilde\tau$
is the optical depth from $\tau$ to today ($\tau_0$). Note that at
late times ($0 \le z \lesssim 10$) we have $\Phi = -\Psi$, since
the anisotropic stress vanishes. As it is the combination
$\Psi'-\Phi'$ which defines the ISW effect, we show on the left
panel of Fig.~\ref{fig:ISWsource} the evolution of $\Psi-\Phi$ for
the same models as in Fig.~\ref{fig:OtherModels}, i.e., Models 1
and 25--28 from Table~\ref{table:Models}. At high redshifts $z
\gtrsim 10^4$, deep in the radiation era, the evolution in the
interacting models is indistinguishable from the non-interacting
model. If we keep all the other parameters fixed to the best-fitting
non-interacting model, then a negative (positive) interaction
leads to faster (slower) decay of the potential $\Psi-\Phi$ around the
time of last scattering; compare the dashed blue (solid black) curves
to the green curve on the left panel of Fig.~\ref{fig:ISWsource}.
This matches to what we already explained about the early ISW
effect: it is more (less) pronounced in the case of negative
(positive) interaction, since the matter-radiation equality
appears later (earlier) than in the non-interacting case. At late
times, $0 \le z \lesssim 10$, the interaction starts to modify the
ISW source \emph{directly}. The negative (positive) interaction
leads to more gradual (steeper) decay of $\Psi-\Phi$, and to a
smaller (an enhanced) late ISW effect.

If we adjust $\omega_c$ and $H_0$ (and hence $\Omega_{de0}$) so
that we obtain the same $z_{\rm eq}$ as in the non-interacting case,
and hence a perfect fit to the acoustic peaks in the data, the
above-described effects on the late ISW effect become even more
pronounced: compare the dot-dashed red (solid cyan) curves to the
green curve on the left panel of Fig.~\ref{fig:ISWsource}. This is
because with negative (positive) $\Gamma$ we need a smaller
(larger) $\Omega_{de0}$, and so the background effect on the
gravitational potential is to reduce (increase) its decay rate.
Now an interesting question arises: how much of the late ISW
effect in the good-fit models comes from the different background,
i.e. different $\Omega_{de0}$ compared to the non-interacting
best-fitting model, and how much comes from the modified perturbation
evolution equations. The right panel of Fig.~\ref{fig:ISWsource}
addresses this question. Blue dashed (black solid) curves show how
the negative interaction Model 27 (positive interaction Model 28)
would behave if we ignored the interaction in the perturbation
equations. Interestingly, we would drastically overestimate the
effect of interaction on the late ISW effect. In the perturbation
equations,
see equations (11--14) or (29--41) in the companion paper \citep*{CompanionPert},
there seems to be a term which partially cancels the
effect of different background evolution. This is easiest to see
in the longitudinal (conformal Newtonian) gauge where $B=0=E$. As
the dark energy perturbations remain subdominant, the interaction
terms in their evolution equations
%in Eqs.~(\ref{eq.delta'de_ourQ}) and (\ref{eq.theta'de_ourQ})
cannot be responsible for the difference.
Moreover, in the CDM velocity equation
%Eq.~(\ref{eq.theta'c_ourQ}),
explicit interaction term is
completely missing. So the only perturbation equation where the
interaction appears to have an effect is the CDM density contrast
equation,
%Eq.~(\ref{eq.delta'c_ourQ}),
which in the longitudinal gauge reads
\be
   \delta'_c = 3\psi' -a\Gamma\phi\,.
\ee
As long as the scale factor is small ($a\ll 1$) the evolution
is like the non-interacting evolution, apart from the different evolution of
the background, but in the later matter
dominated era or dark energy dominated era, when $a$ starts
to approach 1, the interaction starts to modify directly the way the
CDM reflects gravitational potential wells. In this era
there is no anisotropic stress, so $\phi=\psi$,
and let us assume that $\psi>0$. Then due to the
background effect of dark energy, $\psi' < 0$.
Now the effect of the $-a\Gamma\phi$ term is that if $\Gamma < 0$
($>0$), then the CDM density contrast decays slower (faster) than in
the non-interacting case. This in turn feeds back into the
evolution of $\psi$ via an Einstein equation
(the general relativistic Poisson equation,
equation (39) in the companion paper \citep*{CompanionPert}). The effect is
stronger the closer to 1 the scale factor is.

\begin{figure*}
\centering
\includegraphics[angle=-90,width=0.78\textwidth]{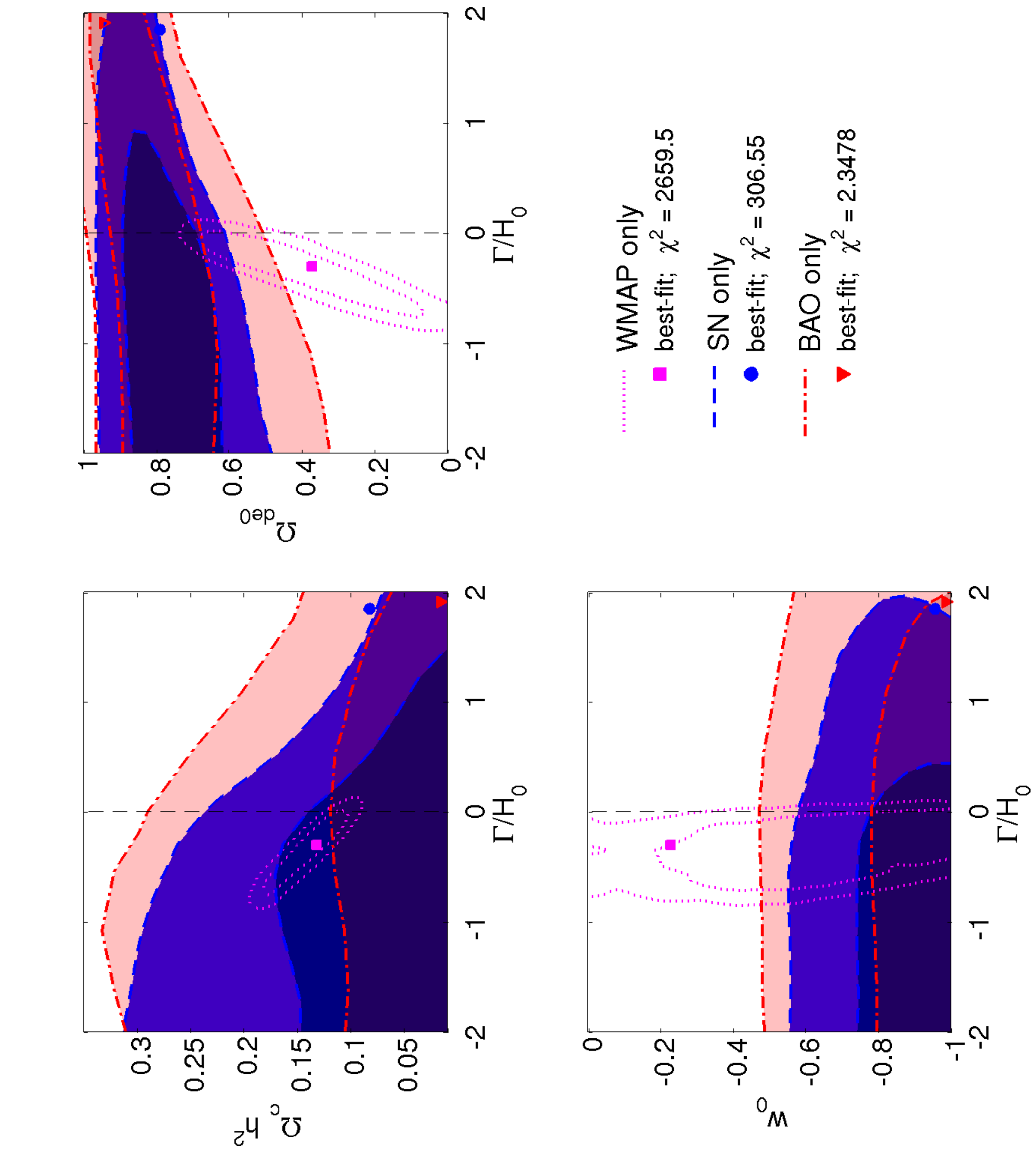}
\caption{2d marginalized likelihoods for the interacting
model with only the WMAP, only the SN, and only the BAO data. The darker blue
or red colours indicate 68\% CL regions while the lighter blue or
red colours indicate 95\% CL regions  with the SN or BAO data,
respectively. The best-fits stand for the best-fitting models in the ranges
shown in this figure. Therefore, here the BAO best-fitting model differs
from the tabulated one which is at $\Gamma/H_0 = 2.92$; see
Table~\ref{table:Models}. \label{fig:LikeIndividual2d}}
\end{figure*}
\begin{figure*}
\centering
\includegraphics[width=0.78\textwidth]{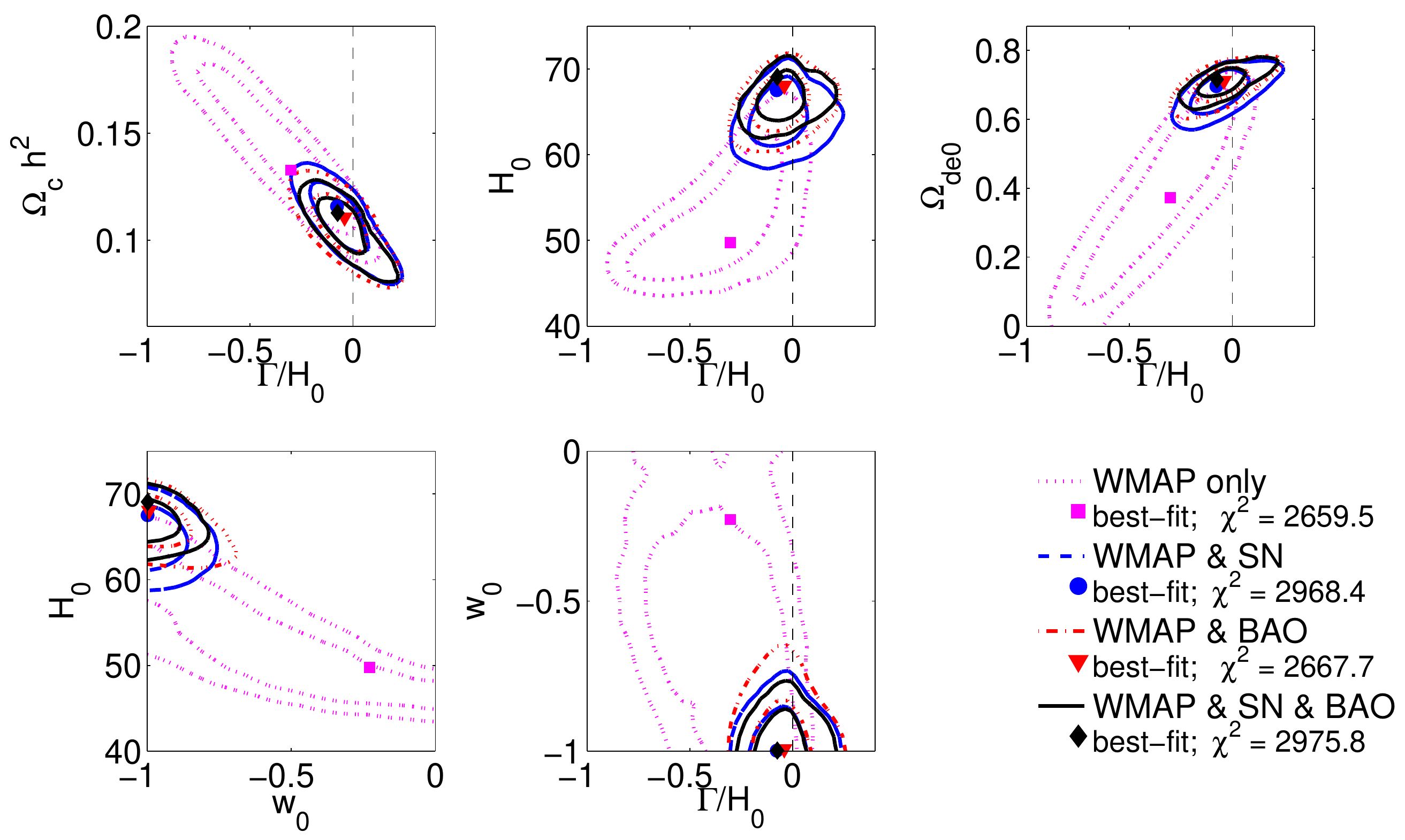}
\caption{2d marginalized likelihoods for the interacting
model with the WMAP, WMAP\&SN, WMAP\&BAO and WMAP\&SN\&BAO data.
\label{fig:Like2d}}
\end{figure*}

\begin{figure*}
\centering
\includegraphics[width=0.7\textwidth]{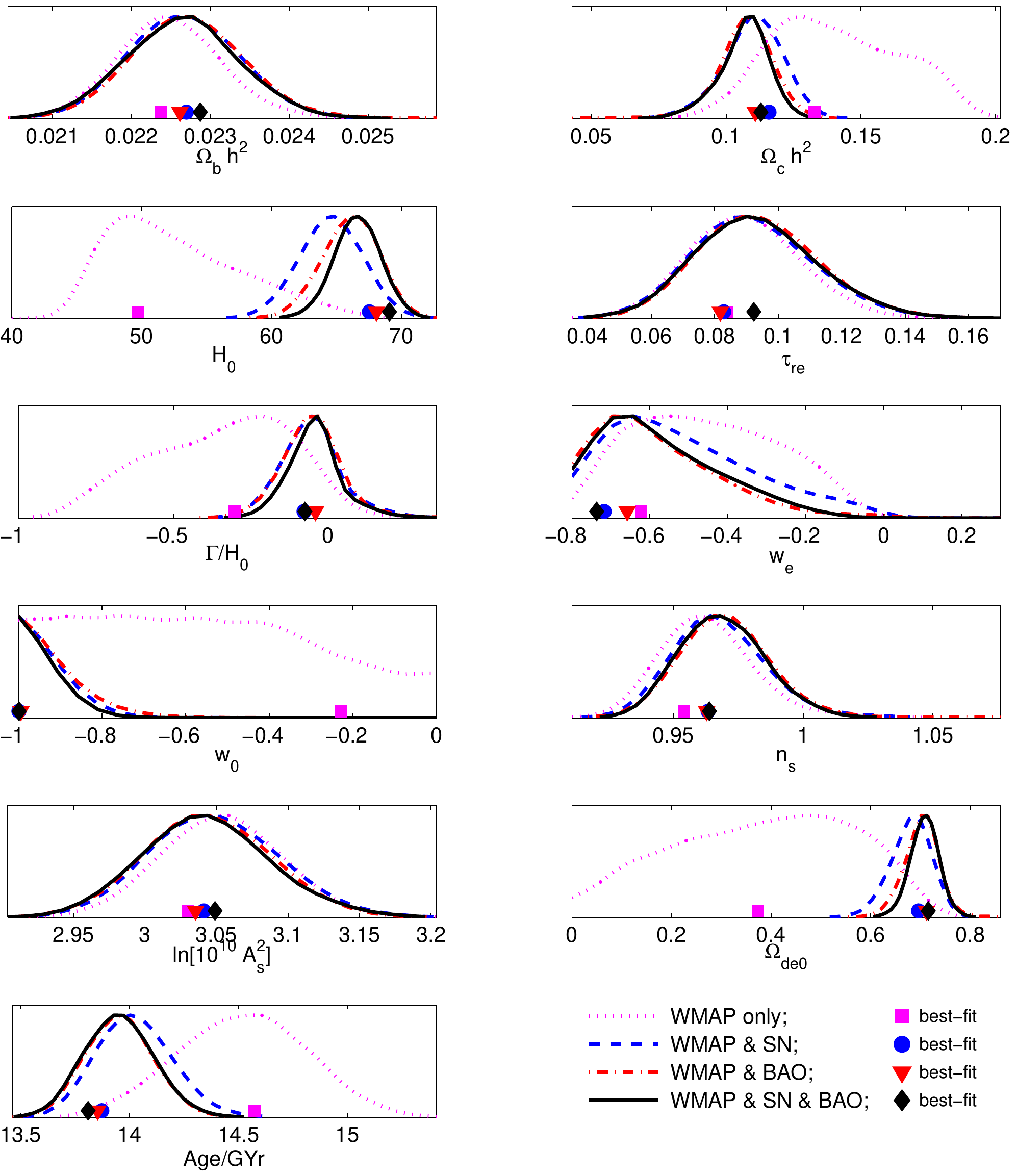}
\caption{1d marginalized likelihoods for the interacting
model with the WMAP, WMAP\&SN, WMAP\&BAO and WMAP\&SN\&BAO data.
\label{fig:Like1d}}
\end{figure*}
\begin{figure*}
\centering
\includegraphics[width=0.7\textwidth]{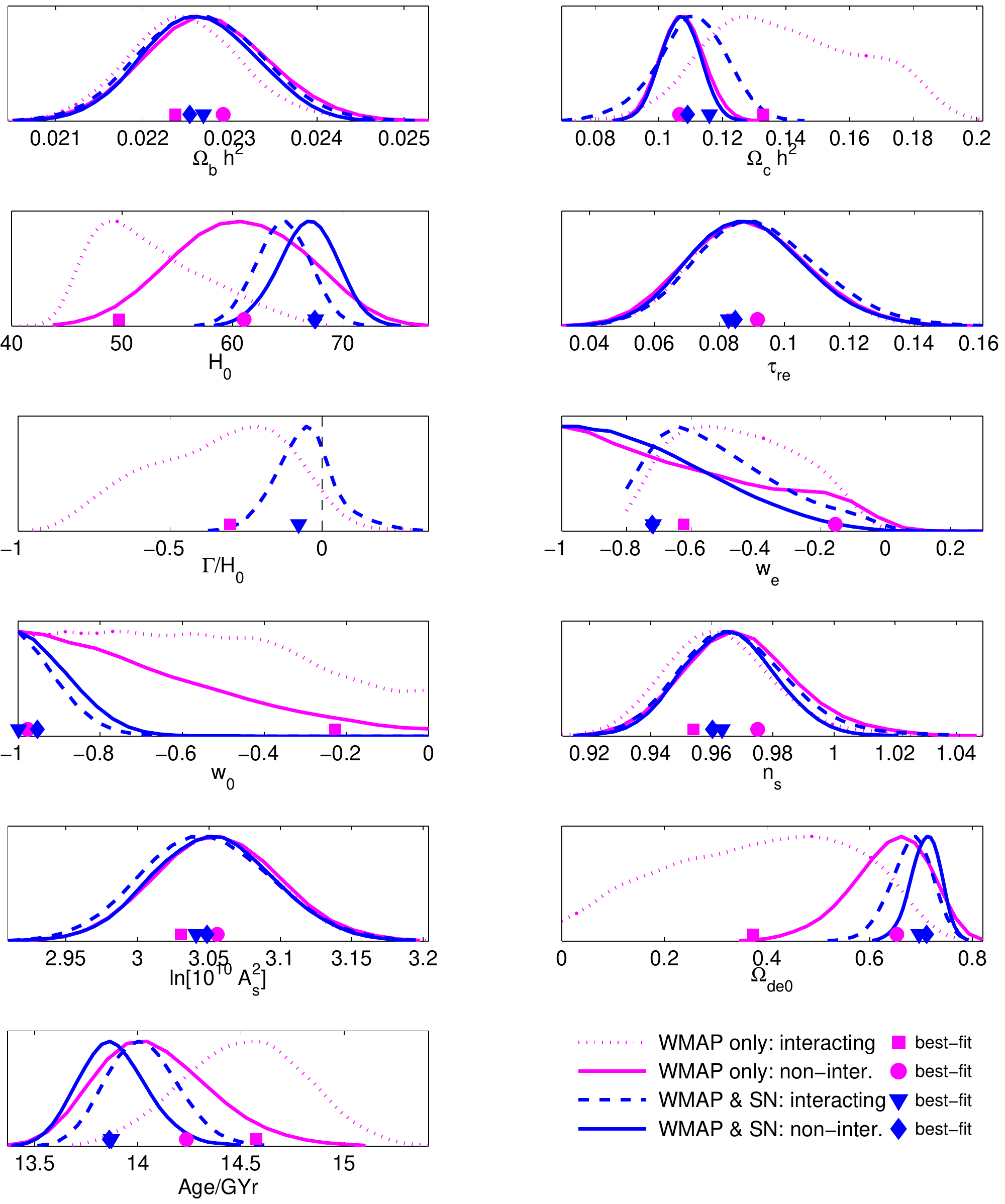}
\caption{Comparison of 1d marginalized likelihoods for the
interacting and non-interacting models with the WMAP and WMAP\&SN
data. The solid lines are for the non-interacting ($\Gamma=0$)
reference model. \label{fig:LikeComp1d}}
\end{figure*}

In the Newtonian gauge we can always neglect the dark energy
perturbations, e.g. the total density contrast is $\delta = \delta\rho_c/\rho_{\rm
tot}$ at late times. However, in synchronous gauge we cannot do
this for the velocity perturbation since  $\theta_c=0$. Indeed, at
late times $\theta_{\rm tot} \approx \theta_{de}$. But the second
Einstein equation is $k^2\eta_{s}' = 4\pi Ga^2(\rho+p)\theta_{\rm
tot}$ -- see e.g. equation (21b) in \cite{Ma:1995ey} -- and this is how the synchronous gauge metric perturbation
$\eta_s$ is calculated in CAMB
%\footnote{http://camb.info/}
\citep{Lewis:1999bs}. So in
synchronous gauge the interaction in the perturbation equations
enters the ISW effect via the interaction in the $\theta_{de}'$
equation, (12) in the companion paper \citep*{CompanionPert}.

Note that it is difficult to go to very large positive
interactions, since the larger the $\Gamma$, the larger $w_e$ one
needs in order to avoid negative $\Omega_{de}$ in the past, i.e.,
to avoid the zero crossing of $\rho_{de}$ which causes the
perturbation equations becoming singular at that moment, as
discussed in Appendix~\ref{app:zerocrossing}.
So we expect the large interaction models to fit the CMB data
as well as the non-interacting model, with
negative $\Gamma$
sightly favoured since this improves the fit in the low multipole
region. Moreover, in marginalized likelihoods the negative
interaction models will be favoured, since the $\rho_{de}$
zero-crossing problem means that there is
much less volume in the allowed parameter space in the positive
$\Gamma$ region than in the negative $\Gamma$ region;
see Appendix~\ref{app:zerocrossing}. There will
be a strong degeneracy between $\omega_c$ and $\Gamma$, $H_0$ and
$\Gamma$, and $\Omega_{de0}$ and $\Gamma$. When adding the other
data (SN or BAO or SN\&BAO) we expect most of the negative
interaction models that fit the CMB alone well, to be excluded due
to their very small  $\Omega_{de}$ today and in the recent past.

\section{Likelihoods}

As predicted, due to the degeneracy between $\omega_c$ and
$\Gamma/H_0$, the CMB data alone do not provide tight
constraints on the interaction. The degeneracy is almost linear,
$\,\omega_c \simeq 0.107 - 0.1\,\Gamma/H_0$, according to
Fig.~\ref{fig:Like2docGamma}, which shows 68\% and 95\% confidence
level (CL) regions with WMAP, WMAP\&ACBAR, and WMAP\&SN\&BAO data.
As explained in the previous section, once $\omega_c$ (and
$H_0$) are adjusted, the interacting model produces completely
indistinguishable CMB angular power spectra at $l \gtrsim 32$.
Therefore, combining WMAP (which reaches up to $l \sim 1000$) with
ACBAR (which reaches $l \sim 2000$) does not help at all. Indeed,
even with the forthcoming Planck data one will not be able to improve
the constraints presented in Fig.~\ref{fig:Like2docGamma}, unless
supplemented with some other non-CMB data. This is because the
only signature from the interaction appears in the ISW region, and
there the accuracy of CMB data is already now cosmic variance
limited.

%\afterpage{\clearpage}
%

%
%
Furthermore, we see from Fig.~\ref{fig:Like2docGamma} that
according to the marginalized likelihoods with the CMB data,
negative interactions are strongly ``favoured'' over positive
interactions. As explained in Appendix~\ref{app:zerocrossing},
this is partially due to a volume effect caused by shrinking of the allowed
$w_e$ direction of parameter space for large positive interaction.
In addition, positive interactions worsen the fit to the
ISW region, while negative interactions improve it. According to
Fig.~\ref{fig:Like2docGamma}, the 95\% CL region of $\Gamma/H_0$
extends from  -0.9 to +0.1. However, it should be noted
that even stronger than -0.9 negative interactions would become
allowed, if we lowered the lower bounds of two of our top-hat
priors: $40 < H_0 < 100$, and $0 < \Omega_{de0} < 1$. This becomes
evident later in the second and third panels of
Fig.~\ref{fig:Like2d}.

Adding SN and BAO data to the analysis leads to a more symmetric
95\% CL region $-0.23 < \Gamma/H_0 < +0.15$, as seen
in Fig.~\ref{fig:Like2docGamma}. In this case, the worse fit to
ISW and the volume effect from $w_e$ in the case of positive
interaction become cancelled by better fits to the
SN and BAO data. In order to gain more insight into this, in some
cases dangerous, competition between the CMB and other data, we
compare in Fig.~\ref{fig:LikeIndividual2d} selected 2d
marginalized posterior likelihoods when using only WMAP or only SN
or only BAO data.

The SN or BAO data alone do not significantly constrain any
parameters of our model other than those shown in
Fig.~\ref{fig:LikeIndividual2d}, i.e., they push $\omega_c$ down
to $\omega_c \lesssim 0.15$, $\Omega_{de0}$ up to $0.65 \lesssim
\Omega_{de0} \lesssim 0.90$, and $w_0$ down to $w_0 \lesssim
-0.75$ at 68\% CL. The non-interacting model ($\Gamma = 0$) is
consistent with all three data sets (WMAP, SN, BAO). Most
interestingly, the non-interacting model sits in the intersection of all
three data sets so that there is no tension between them.
As already noticed, there is a tension between CMB and
SN or CMB and BAO in the negatively interacting models. This
tension is most pronounced in the top right panel
($\Gamma,\Omega_{de0}$) of Fig.~\ref{fig:LikeIndividual2d}.
However, the situation is not too bad since there is plenty of
parameter space volume in the intersection of 95\% CL regions of
WMAP and SN or WMAP and BAO.
%We just need to keep in mind that
%when combining any CMB data set with SN or BAO there will be some
%competition between the slightly improved CMB fit and the slightly
%worsened SN or BAO fit in the negative interaction models.
%Therefore,
Although the \emph{SN or BAO data do not put any direct
constraints on the interaction} (even $|\Gamma/H_0|\sim 3$ fits
them well) the net effect of combining CMB with SN or BAO data is
to force $\Omega_{de0} \gtrsim 0.65$ and hence to cut away the
large negative interaction models, leaving an almost symmetric region
around $\Gamma=0$. This effect is seen in the top left panel of
Fig.~\ref{fig:Like2d}, where we show 2d marginalized likelihoods
from our MCMC runs for the interacting model with the WMAP data
alone, the WMAP\&SN, the WMAP\&BAO, and the WMAP\&SN\&BAO data.
The SN and BAO data are rather consistent with each other when
constraining the interacting model. Therefore combining WMAP with
either or both of them leads to very similar constraints, as seen
in Fig.~\ref{fig:Like2d}.

In Fig.~\ref{fig:Like1d}
%in Appendix~\ref{App:1d},
we show the 1d
marginalized likelihoods for all of the primary MCMC parameters of
our model and for two derived parameters: $\Omega_{de0}$ and the
age of the universe. We note that WMAP combined with the BAO data
prefers slightly larger today's Hubble parameter $H_0$ and dark
energy density $\Omega_{de0}$ than with the SN data.
The CMB data alone favour negative interactions and thus require
small $H_0$ and  $\Omega_{de0}$. This corresponds to a very old
universe as seen in the last panel of  Fig.~\ref{fig:Like1d}.

As mentioned above,
from the 1d plot for $\Gamma/H_0$ in Fig.~\ref{fig:Like1d},
the negative interaction seems more probable than the positive
interaction. Indeed with the WMAP data (WMAP\&SN\&BAO data) 96.8\%
(77.6\%) of models in our Markov Chains have a negative $\Gamma$,
which means energy transfer from dark energy to dark matter.
However, we should be cautious in claiming that the WMAP data
(WMAP\&SN\&BAO data) favour energy transfer from dark energy to CDM at
96.8\% CL (77.6\% CL). It should again be stressed that this is
partially the volume effect from the $w_e$ direction of the parameter space.
%%%%%%%%%%%%%%
Indeed, recently~\citet{Pereira:2008at} claimed
that with 93\% probability the data -- which in \citet{Pereira:2008at} were the
background-based data only -- favour decay of dark
matter to dark energy. Now we have about the same probability in
favour of energy transfer from the dark energy to dark matter. The
message here is that, in addition to the volume effect,
these claims are highly model dependent:
\citet{Pereira:2008at} studied an interaction proportional
to $\rho_{de}$ whereas our interaction is proportional to $\rho_c$.
%%%%%

We summarize here our most stringent results for the interacting
model by giving minimal 95\% intervals \citep{Hamann:2007pi} from our
MCMC run with WMAP\&SN\&BAO data: $\omega_b \in (0.0212,\,
0.0241)$, $\omega_c \in (0.859, \,  0.125)$, $H_0 \in (63,\, 70)$,
$\tau \in (0.057,\, 0.133)$, $\Gamma/H_0 \in (-0.23,\, +0.15)$,
$w_e \in (-0.80,\, -0.19)$, $w_0 \in (-1.00,\, -0.63)$, $n_S \in
(0.937,\, 1.002)$, $\ln(10^{10}A_S^2) \in (2.95,\, 3.14)$,
$\Omega_{de0} \in (0.648,\,  0.767)$, Age$\,\in (13.6,\,
14.3)\,$Gyr.

\begin{figure}
\centering
\includegraphics[width=0.47\textwidth]{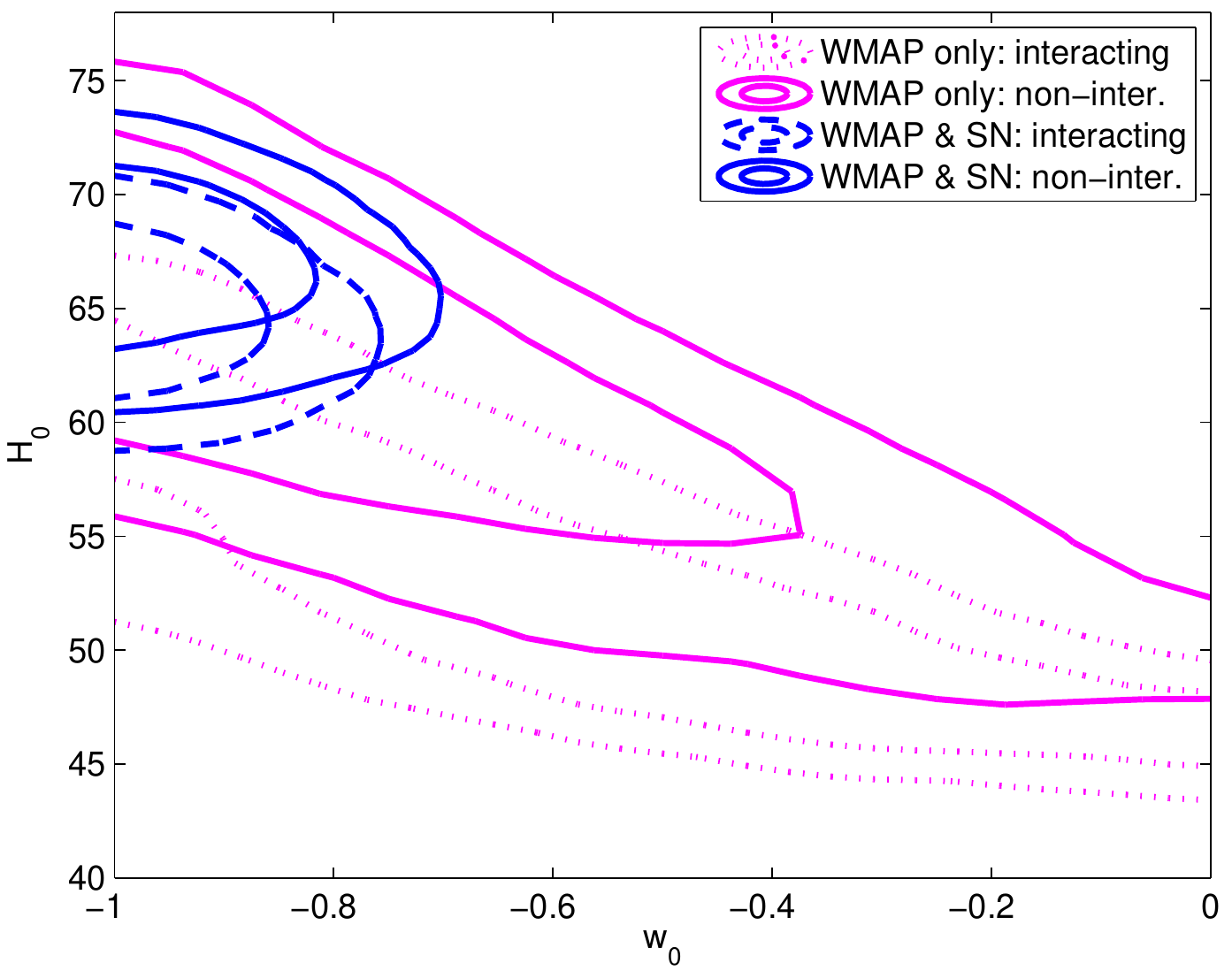}
\caption{Comparison of a 2d marginalized likelihood for the
interacting and non-interacting models with the WMAP and WMAP\&SN
data. Note the degeneracy between $H_0$ and $w_0$. The solid lines are for
the non-interacting ($\Gamma=0$) reference model.
\label{fig:LikeComp2d}}
\end{figure}

In Fig.~\ref{fig:LikeComp1d}
%in Appendix~\ref{App:1d},
we compare the 1d marginalized likelihoods of the interacting model to the
non-interacting reference model ($\Gamma=0$). The key differences are:
the interacting model leads to broader distributions of
$\omega_c$, $H_0$ and $\Omega_{de0}$. This is due to the
degeneracy between $\omega_c$ (or $H_0$ or $\Omega_{de0}$) and
$\Gamma$. Moreover, smaller $H_0$ is favoured by the interacting
model. However, note again that this is partially a parameter
space volume effect, since in our MCMC chains there are many more
negative interaction models than positive interaction models, and
the good-fit negative interaction models have small $H_0$. Finally
we note that $H_0$, both in the interacting and non-interacting cases,
is smaller than in the $\Lambda$CDM model, where a typical result of a
likelihood scan peaks around $H_0 = 72\,$km$\,$s$^{-1}$Mpc$^{-1}$. This is because
we let $w_0$ (and $w_a$) vary, and there is a strong degeneracy
(even in the non-interacting case) between $H_0$ and $w_0$ (or $w_{de}$).
From Fig.~\ref{fig:LikeComp2d} we see that a $w_0 = -1$ model
prefers largest $H_0$ (and if we allowed $w_0<-1$, even larger
values of $H_0$ would be favoured).

\begin{figure*}
\centering
\includegraphics[width=\textwidth]{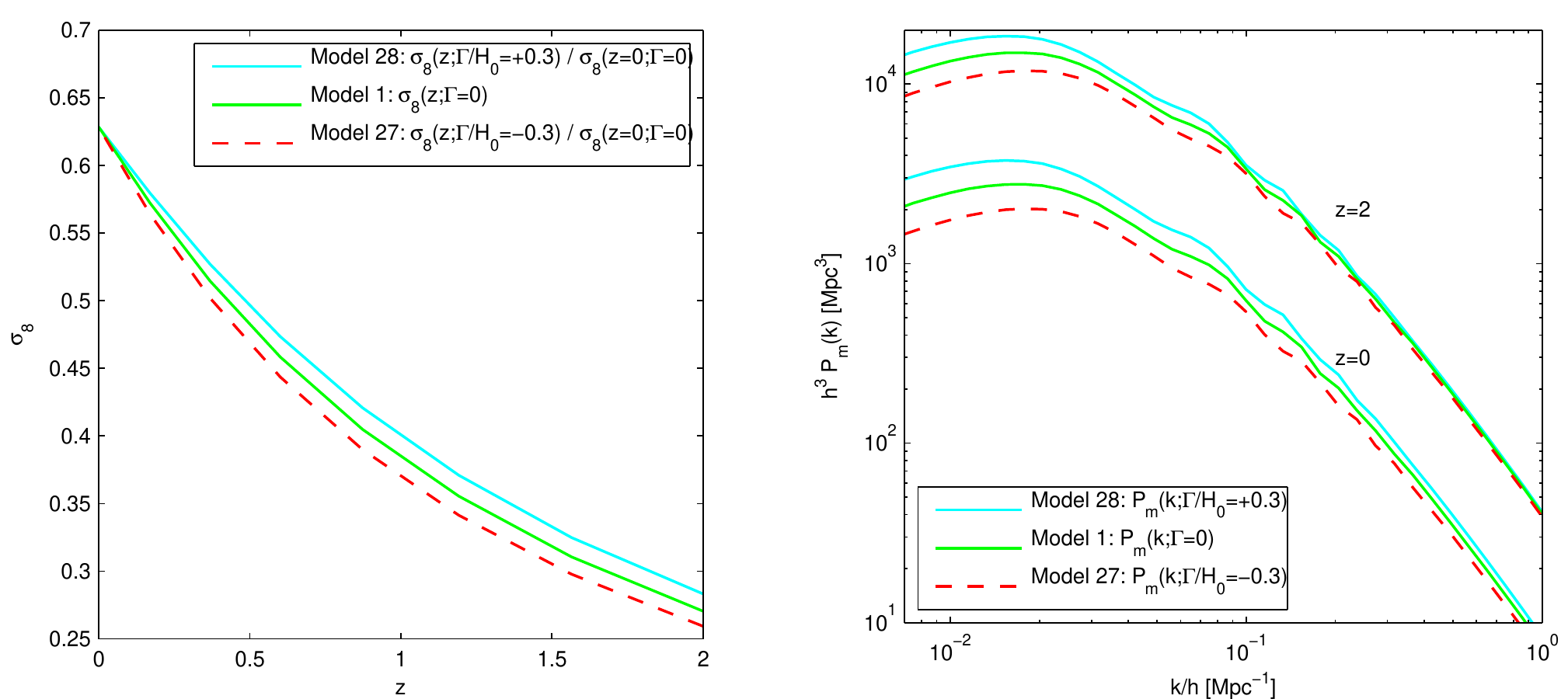}
\caption{Redshift evolution of $\sigma_8$ and matter power
spectrum. \label{fig:MatterPower}}
\end{figure*}

\section{Additional constraints on the model}

In the two previous sections we have seen that any CMB data
alone fail to constrain the interacting model, in particular
with negative interaction, but the constraints from the SN or BAO
data for the background dark energy density in the recent past
cut out the strong interaction models with $\Gamma<0$,
as these models have very small $\Omega_{de}$.
In this section we briefly discuss  some other
data that could be used.

Most interestingly, as the interaction seems to affect the ISW
effect only, accurate CMB -- large-scale structure (LSS)
cross-correlation data on the ISW effect (combined
with other data sets such as CMB, SN and BAO) may turn out to
set the most stringent constraints on the interaction. Firstly, the
CMB--LSS cross--correlation data probe the evolution of
$\Omega_{de}$ over time in the recent past. Secondly, as the
interaction affects not only the background evolution, but
directly the evolution of perturbations (in particular, the way that
the evolution of the cold dark matter perturbation and gravitational
potential are linked to each other) at redshifts probed by the ISW
data, the ISW data may see the effects from the interaction most
directly. We are working on this issue. While we were finalizing this paper,
a work on other type
of dark sector interaction appeared also suggesting that the
interaction could be detected via its effects on the ISW signal
\citep{He:2009pd}. Earlier \cite{Lee:2006za} discussed the
modified ISW effect in an interacting quintessence model.
However, \emph{they fixed all the other parameters} to (or near to) the
best-fitting to WMAP Lambda CDM model. Therefore, although mentioning the
degeneracy  between $\Omega_{de0}$ (or $\omega_c$) and the interaction rate elsewhere
in their paper, they missed the degeneracy when finding
constraints on the interaction rate with WMAP data, ending up with artificially
tight constraints on the interaction with CMB data alone. This can be
seen, for example, in the third panel of our Fig.~\ref{fig:Like2d}. Had we
fixed $\Omega_{de0}$ to $0.76$, as \cite{Lee:2006za} did, our constraint
on the interaction rate would be misleadingly tight:
$-0.03 \lesssim \Gamma/H_0 \lesssim +0.005$.
It should be noted that although adding SN or BAO or SN\&BAO into the analysis
forces $\Omega_{de0}$ (and $\omega_c$ and $H_0$) closer to the
best-fitting (to WMAP) Lambda CDM model and hence leads to a similar kind of effect
as fixing the parameters, letting the other parameters freely vary and 
consequently \emph{taking properly into account the degeneracy} gives much looser constraint:
$-0.23 < \Gamma/H_0 < +0.15$ with WMAP\&SN\&BAO. So fixing the other parameters
would misleadingly give more than an order of magnitude smaller 95\% interval in our case.

Other interesting data come form the galaxy-galaxy power spectrum which
is a probe of the underlying theoretical  matter power spectrum.
However, we have decided not to use these data here, since the
exact relation between the observed galaxy-galaxy power spectrum
and matter power spectrum is not known (due to redshift space
distortions, non-linearities, etc) and moreover we calculate the
perturbation evolution in the linear regime only. Probably the
interaction would affect non-linear structure formation
\citep{Baldi:2008ay,Baldi:2009yv}.
So at
the moment it does not make sense to fit the shape of the (linear) matter power
to the data. For similar reasons we do not consider weak-lensing data.

As the strongly interacting Models 27 and 28 in
Table~\ref{table:Models} have very different recent-time evolution
of matter and dark energy densities, the overall amplitude of the
matter power is affected. This results in different $\sigma_8$
(root mean square mass fluctuation on $8h^{-1}\,$Mpc scale) than
in the non-interacting model. We find that today $\sigma_8(z=0)=$
0.598, 0.628, 0.654 for Model 27 ($\Gamma/H_0=-0.3$), Model 1
($\Gamma=0$), and Model 28 ($\Gamma/H_0=+0.3$), respectively.
Therefore, we predict that $\sigma_8$ measurements could rule out
large negative interactions as they may lead to too small
$\sigma_8$. On the left panel of Fig.~\ref{fig:MatterPower} we
show the redshift evolution of $\sigma_8$. We have normalized the
curves for the interacting models to fit today's $\sigma_8$ of the
non-interacting model. For example, for the $\Gamma/H_0=-0.3$ model
$\sigma_8(z=0)$ is smaller than in the non-interacting case, and
towards the past it decreases even faster than in the non-interacting
case. On the right panel of Fig.~\ref{fig:MatterPower} we compare
matter power spectra at redshifts $z=0$ and $z=2$ in the strongly
interacting and non-interacting models. Small-scale (large $k$)
shape is unaffected by the interaction, whereas large-scale (small
$k$) shape (and amplitude) are affected significantly.
Marginalizing analytically over the galaxy bias, the SDSS DR4
Luminous Red Galaxies sample \citep{Tegmark:2003uf,Tegmark:2006az}
gives the best $\chi^2$ for the positive interaction
and worst $\chi^2$ for the negative interaction, while the $\chi^2$
of the non-interacting model falls between these two.\
As the effect of adding the LSS data to the analysis seems to be
very similar to the effect of adding BAO or SN,
we expect that adding the LSS data would give more weight to the positive interactions and hence lead to even more symmetric and tighter probability distribution about
$\Gamma = 0$ than with the CMB\&SN\&BAO data.

%%%%%%%%%%%%%%%%%%%%%%%%%%%%%%%%%%%%%%%%%%%%%%%%%%%%%%%%%%%
%Model  1 Uncoupled MPK chi2 = 53.07408   sigma_8 = 0.628 %
%Model 27 Neg Gamma MPK chi2 = 63.58498   sigma_8 = 0.598 %
%Model 28 Pos Gamma MPK chi2 = 44.08401   sigma_8 = 0.654 %
%%%%%%%%%%%%%%%%%%%%%%%%%%%%%%%%%%%%%%%%%%%%%%%%%%%%%%%%%%%

\section{Conclusion}

In the companion paper \citep*{CompanionPert} we have presented, for the first time,
a \emph{systematic} derivation of initial conditions for perturbations in
interacting dark matter -- dark energy fluid models deep in the
radiation dominated era. These initial conditions are essential
for studying the further evolution of perturbations up to today's
observables.
%In this paper we have implemented the perturbation
%equations and adiabatic initial conditions into a Boltzmann integrator CAMB
%which solves the multipole hierarchy and
%produces the theoretical prediction for the CMB temperature and
%polarization angular power as well as the matter power spectrum
%etc.
We have focused on the interaction $Q_c^\mu =- \Gamma\rho_c
(1+\delta_c) u_c^\mu $, where $\Gamma$ is a constant which has the
same dimension as the Hubble parameter $H$; see
Eqs.~(\ref{eq.cont_rhoc}) and (\ref{eq.cont_rhode}).
%In line with
%a previous result for a non-interacting quintessence dark energy
%in \citep{Doran:2003xq}, we find that, in our interacting model,
%requiring adiabaticity between all the other constituents
%(photons, neutrinos, baryons, and cold dark matter) leads
%incidentally also the dark energy to become adiabatic, if its
%early time equation of state parameter is $w_e < -1$ or $-2/3 <
%w_e \le 1/3$.
In our previous work \citep{Valiviita:2008iv} we
showed that if the equation of state parameter for dark energy is
$-1 < w_{de} < -4/5$ in the radiation or matter dominated era, the
model suffers from a serious non-adiabatic instability.
%In this
%paper the systematic derivation of initial conditions confirms the
%result.
However, in this paper (and in the companion paper \citep*{CompanionPert}) we have shown that
the instability can easily be avoided, if we allow for suitably
time-varying dark
energy equation of state. Our worked out example is for the
parametrization $w_{de} = w_0a + w_e (1-a)$. With this
parametrization, viable cosmologies (in the interacting model)
result whenever $w_0$ is close to $-1$ and $w_e < -1$ or
$-4/5 \le w_e \le 1/3$, as long as $w_0+1$ and $w_e+1$
have the same sign.

We have implemented into a publicly available Boltzmann
integrator, CAMB, the background equations, and first order
(linearized) perturbation equations, as well as the adiabatic
initial conditions for the interacting model with
time-varying equation of state parameter. We have
performed full Monte Carlo Markov Chain likelihood scans for this
model as well as for the non-interacting ($\Gamma$=0) model for a
reference, with various combinations of publicly available data
sets (WMAP, WMAP\&ACBAR, SN, BAO, WMAP\&SN, WMAP\&BAO,
WMAP\&SN\&BAO). To avoid ending up with too complicated shape of
parameter space we have focused on non-phantom models,
$-1 < w_{de} < 1/3$.
(In addition, we consider phantom models to be unphysical.)

The main result is that there is a degeneracy between
the interaction rate and today's dark energy (or dark matter)
density in light of CMB data. Therefore, CMB data alone cannot
rule out large interaction rates, not even Planck, since the
high-multipole part of the CMB angular power spectra are
totally indistinguishable from
the non-interacting case. The only signal remaining
from a large interaction rate
would be a modified integrated Sachs-Wolfe effect, which makes using
the ISW data an appealing line of future work. In this paper we have broken the
degeneracy by Supernovae data and by baryon acoustic oscillation
data, finding that the CMB\&SN\&BAO data constrain
the interaction rate to about 20\% of the expansion rate
of the Universe.

\vspace{2mm} {\bf Acknowledgments:} JV and RM are supported by
STFC. During this work JV received support also from the Academy
of Finland.
We thank Daniele Bertacca for comments, and
acknowledge use of CosmoMC and CAMB.
The MCMC analysis was mainly conducted in cooperation
with SGI/Intel utilizing the Altix 3700 supercomputer
at the UK-CCC facility COSMOS.

%\clearpage
\appendix

\section{Technical details }
\label{sec:technicaldetails}
\subsection{The code and modified sound horizon \label{app:code}}

We have modified publicly available CosmoMC
%\footnote{http://cosmologist.info/cosmomc}
\citep{Lewis:2002ah,cosmomc_notes} and CAMB
%\footnote{http://camb.info/}
\citep{Lewis:1999bs} for this
study. Into CAMB we have implemented the interacting evolution
equations for the background (\ref{eq.cont_rhoc}) and
(\ref{eq.cont_rhode}), as well as the interacting perturbation
evolution for dark energy, equations (11--14)
from the companion paper \citep*{CompanionPert},
%Eqs.~(\ref{eq.delta'de_ourQ}) and (\ref{eq.theta'de_ourQ}),
in the synchronous gauge: $B=\phi=0$,
$\psi = \eta_s$, and $E = -k^{-2}(6\eta_s+h_s)/2$, with $\eta_s$
and $h_s$ representing the synchronous gauge metric perturbations.
In the synchronous gauge, the perturbed CDM equations of motion
%Eqs.~(\ref{eq.delta'c_ourQ}) and (\ref{eq.theta'c_ourQ}),
appear to look the same in the interacting and non-interacting cases,
since $\phi=0$. However, it should be noted that as the background
evolves differently, it affects the evolution of $h_s$, and hence
$\delta_c$. We have implemented the initial conditions for
perturbations deep in the radiation dominated era, specified in
the companion paper \citep*{CompanionPert}.

In order to use the BAO data in the interacting model an
additional modification to CAMB is necessary. In the standard
version of CAMB/CosmoMC, the sound horizon is calculated by
numerically integrating up to last scattering $a_\ast$ \be
r_s(a_{\ast}) = \int_0^{\tau_\ast} c_s(\tau) d\tau =
\int_0^{a_\ast} c_s(a)/a' da\,, \ee where $c_s$ is the sound speed
in the photon-baryon fluid and $a_\ast$ is very accurately calculated
from a fitting formula \citep{Hu:1995en} which is valid if the matter scalings are
the standard ones, $\rho_b \propto a^{-3}$ and $\rho_c \propto
a^{-3}$. Firstly we need a different formula, as for BAO we want to
integrate up to decoupling $a_{\rm dec}$, for which one has another
fitting formula \citep{Hu:1995en}. Also this formula is valid only if $\rho_b
\propto a^{-3}$ and $\rho_c \propto a^{-3}$. However, we know that
this is not true for our case. Therefore we find $a_{\rm dec}$
numerically, and then calculate numerically
\be
r_s(a_{\rm dec}) =
\int_0^{a_{\rm dec}} c_s(a)/a'\, da\,.
\label{eq:rsdec}
\ee
The defining equation of decoupling is \citep{Hu:1995en}
\be
-\int_{\tau_0}^{\tilde\tau}
R(\tau) \times \mbox{opacity}(\tau)\, d\tau = 1\,,
\label{eq:decdefinition}
\ee
where $R = \frac{3}{4}  \rho_b / \rho_\gamma$.
%, and the opacity is conventionally denoted by $\tau'$, but as we use $\tau$ to denote the conformal time we cannot follow the convention.
We numerically follow this integral from today toward past times
$\tilde\tau$ until the value $1$ is reached. Then we record the
value of the scale factor at this moment, name it $a_{\rm dec}$ and
convert to a redshift $\tilde z_{\rm dec} = 1/a_{\rm dec} -1$.
Finally, to match the definitions in \citet{Eisenstein:2006nk}
and \citet{Percival:2007yw} (from where we take the BAO data)
we multiply the result by their ``phenomenological'' factor $0.96$;
$z_{\rm dec} = 0.96
\tilde z_{\rm dec}$. We have verified that in the non-interacting
case these definitions and our numerical routines lead to the same
$z_{\rm dec}$ and $r_s(a_{\rm dec})$ as those given as a test case
in \citet{Percival:2007yw}.

\begin{figure*}
\centering
\includegraphics[width=0.95\textwidth]{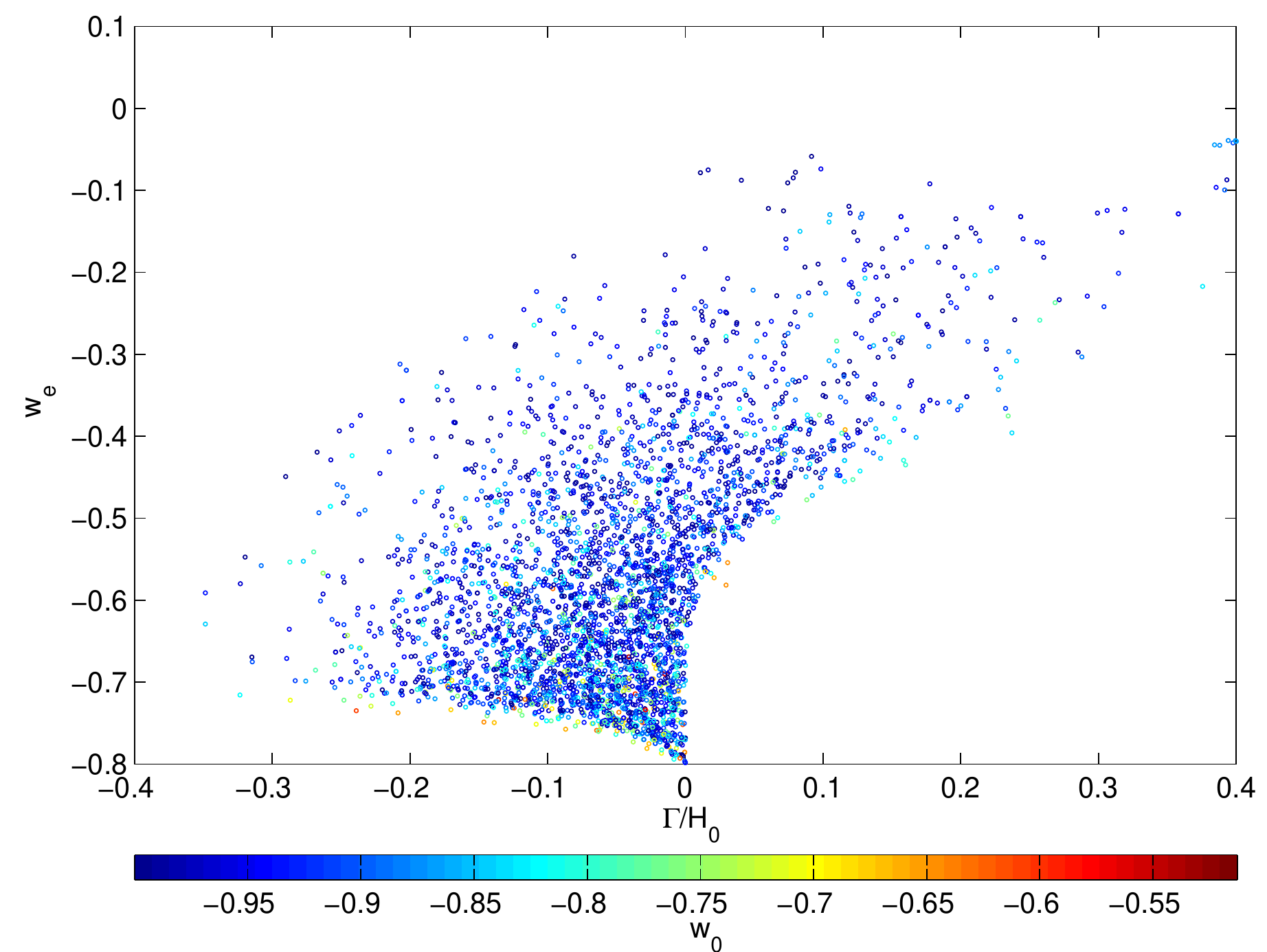}
\caption{The $w_e$ volume effect. The circles show
samples from our MCMC run with the WMAP\&BAO data. The colour
scale indicates the value of $w_0$ for each sample.
\label{fig:VolumeEffect}}
\end{figure*}

\subsection{The parameters and their prior ranges \label{app:ranges}}

\begin{table*}
 \begin{tabular}{|l|l|l|}
 \hline
 symbol & explanation & defining equation / reference\\
 \hline
subscript $c$ & cold dark matter & \\
subscript $de$ & dark energy & \\
subscript $b$ & baryons & \\
subscript $0$ & variable evaluated today & \\
$\tau$ & conformal time &  $d\tau = a^{-1} dt$, where $a$ is the scale factor of the Universe \\
${}'$ & conformal time derivative & E.g. $a' = \frac{da}{d\tau}$ \\
$H$ & Hubble parameter, i.e., the expansion rate of the Universe & $H= \frac{da}{dt}/a$\\
$\h$  & conformal Hubble parameter &  $\h = a'/a = aH$\\
$\omega_b$ & physical density parameter of baryons & $\omega_b = \Omega_{b0} h^2$\\
$\omega_c$ & physical density parameter of cold dark matter & $\omega_c = \Omega_{c0} h^2$\\
$\rho_{\rm tot}$ & total density & sum of energy densities of all constituents of the Univ. \\
$n_S$  & scalar spectral index & \\
$A_S$ & amplitude of the primordial curvature perturbation power & $\mathcal{P}_{\mathcal{R}} (k) = A_S (k/k_0)^{n_S - 1}$, where  $k_0 = 0.05\;$Mpc$^{-1}$\\
$\tau_{re}$ & optical depth to reionization & \\
$D_V$ & dilation scale & equation (2) of \citet{Eisenstein:2005su}\\
$D_A$ & angular diameter distance & \\
$z_{\rm dec}$ & redshift of decoupling & Eq.~(\ref{eq:decdefinition}) and explanation after that: $z_{\rm dec} = 0.96\tilde z_{\rm dec}$ \\
$z_\ast$ & redshift of last scattering & \\
$r_s$ & sound horizon a decoupling & Eq.~(\ref{eq:rsdec}) \\
$\phi$, $\psi$, $B$, $E$ & scalar metric perturbation variables & equation (6) of   \citet{CompanionPert}\\
$\Phi$, $\Psi$& scalar gauge invariant metric perturbations & equations (6) and (25) of \citet{CompanionPert}\\
$\eta_s$, $h_s$ & synchronous gauge metric perturbations & these are called $\eta$ and $h$ in \citet{Ma:1995ey}\\
$\theta_c$ & cold dark matter velocity perturbation & see e.g. \citep{Valiviita:2008iv,Ma:1995ey} \\
$\theta_{de}$ & dark energy velocity perturbation & \\
\hline
 \end{tabular}
\caption{List of selected symbols used in this paper, with their physical meaning and defining equation or a reference.\label{table:symbols}}
\end{table*}
In the interacting model we have 9 primary MCMC parameters which
we vary over wide ranges with uniform (flat) prior over their range:
the physical baryon density today
$\omega_b = h^2 \Omega_{b0}$, the physical cold dark matter density
today $\omega_c = h^2 \Omega_{c0}$, the Hubble parameter today $H_0$,
optical depth to reionization $\tau_{re}$, the interaction $\Gamma$ in
units of today's Hubble parameter, the early dark energy equation
of state parameter $w_e$, the dark energy equation of state
parameter today $w_0$, scalar spectral index $n_S$ of the
primordial perturbations, and the amplitude of primordial
perturbations $A_S$ [we use $\log_e(10^{10}A_S^2)$ as in the
standard CosmoMC]. In the non-interacting reference model we have
8 free parameters as we keep $\Gamma$ fixed to zero.

We exclude phantom models ($w_{de}<-1$) as unphysical,
and we focus on the following ranges
\be
-1  <  w_0 < 0\,, \quad -1 < w_e < 0.3\,.
\ee
It should be noted that the CMB data
actually exclude all blow-up models, i.e., the interacting models
with  $-1 < w_e < -0.8$. Keeping this in mind, our motivation to
drop phantom models from the analysis becomes reinforced: it
would make the posterior of $w_e$ consist of two separate
intervals, and hence the posterior likelihood of $w_e$ would have
two peaks, one with $w_e < -1$ and another one somewhere in the
range $-4/5 \le w_e < 1/3$. As the MCMC technique is particularly
ill-suited for multiple peaked posteriors, allowing for phantom
models would ruin the analysis. Moreover, we should demand $w_e+1$
and $w_0+1$ to have the same sign, since otherwise there would be a
time $\tau_{\mathrm{oc}}$ in the past where
$w_{de}(a_{\mathrm{oc}})=-1$. Such a
'minus one' crossing causes severe problems since the dark energy
perturbation
equations contain terms proportional to $1/(1+w_{de})$,
see the companion paper \citep*{CompanionPert}.
%, for more details see, e.g., \citep{xxxxx}.
The requirement ${\rm sign}(w_e+1)={\rm sign}(w_0+1)$
would further complicate the shape of the parameter space,
if we allowed for $w_e < -1$ or $w_0 < -1$.

As discussed in \citet{Valiviita:2008iv}, $|\Gamma/H_0| \gg 1$
would conflict with the data. We find that a reasonable range
is $\Gamma/H_0 \in (-4,\, +4)$. The ranges of other remaining
primary MCMC parameters are: $\omega_b \in (0.005,\, 0.1)$,
$\omega_c \in (0.01,\, 0.99)$, $H_0 \in (40,\, 100)$, $\tau \in
(0.01,\, 0.4)$, $n_S \in (0.5,\, 1.5)$, $\ln(10^{10}A_S^2) \in
(2.7,\, 4.0)$. Finally, we have an amplitude of the SZ template in
CosmoMC, $A_{SZ} \in (0,\ 2)$, which is used when adding the SZ
templates to the high-$l$ part of the TT angular power spectrum.
We also restrict the analysis to positive dark energy densities
today, i.e., apply a top-hat prior $\Omega_{de0} > 0$. As we study
spatially flat models, the matter density today is $\Omega_{m0} =
1 - \Omega_{de0}$.

In Table \ref{table:symbols} we list some of the above-mentioned
parameters as well as other symbols used in this paper.

\begin{table*}
\caption{The $\chi^2$s and parameters of the best-fitting model from
each of our 6 MCMC runs for the non-interacting model ($\Gamma=0$) and from our
6 MCMC runs for the interacting model ($\Gamma\neq 0$).
The rows ``neg. $\Gamma$''
and ``pos. $\Gamma$'' show the models with largest $|\Gamma|$
(found in our MCMC chains) such that $\chi^2$ is within 4 ($\sim
2\sigma$ if the likelihood was Gaussian) from the corresponding best-fitting
model. \label{table:Models}}
%\scriptsize
\begin{tabular}{|rr|l||r||r|r|r|r|r|r}\hline
\multicolumn{2}{|l|}{Model} & Data & $\chi^2$ &
$\!\chi^2_{\!\mathrm{WMAP}}$ & $\chi^2_{l<32}$ & $\chi^{2\mathrm{TT}}$ & $\chi^{2\mathrm{TT}}_{l<32}$ & $\chi^2_{\mathrm{SN}}$ & $\!\chi^2_{\!\mathrm{BAO}}$\\ \hline
%\hline
%WMAP
%%%%%%%%%%%%%%%%%%%%%%%
 1 & best $\Gamma\!=\!0$  & WMAP          & 2660.4 & 2660.4 & 1221.5 & 1023.5 & -11.3 & 311.4 & 22.5 \\
 2 & best $\Gamma\!\neq\! 0$    & WMAP          & 2659.5 & 2659.5 & 1221.5 & 1022.8 & -11.6 & 368.2 & 63.7 \\
 3 & neg.\ $\Gamma$      & WMAP          & 2662.6 & 2662.6 & 1222.0 & 1024.9 & -11.1 & 389.4 & 81.8 \\
 4 & pos.\ $\Gamma$      & WMAP          & 2663.4 & 2663.4 & 1222.2 & 1024.9 & -11.0 & 316.1 & 27.6 \\
%%%%%%%%%%%%%%%%%%%%%%
%\hline
%SN
%%%%%%%%%%%%%%%%%%%%%%%
 5 & best $\Gamma\!=\!0$  & SN           & 307.6 & -- & -- & -- & -- & 307.6 & 1260 \\
 6 & best $\Gamma\!\neq\! 0$    & SN           & 306.5 & -- & -- & -- & -- & 306.5 & 146.7 \\
%%%%%%%%%%%%%%%%%%%%%%
%\hline
%BAO
%%%%%%%%%%%%%%%%%%%%%%%
 9 & best $\Gamma\!=\!0$  & BAO          & 3.3 & -- & -- & -- & -- & 334.9 & 3.3 \\
10 & best $\Gamma\!\neq\! 0$    & BAO          & 2.1 & -- & -- & -- & -- & 363.7 & 2.1 \\
%%%%%%%%%%%%%%%%%%%%%%
%\hline
%WMAP and SN
%%%%%%%%%%%%%%%%%%%%%%%
13 & best $\Gamma\!=\!0$  & WMAP\&SN      & 2968.7 & 2660.7 & 1222.2 & 1023.6 & -10.5 & 308.2 & 7.8\\
14 & best $\Gamma\!\neq\! 0$    & WMAP\&SN      & 2968.4 & 2660.3 & 1221.0 & 1024.1 & -11.9 & 308.1 & 7.8\\
15 & neg.\ $\Gamma$      & WMAP\&SN      & 2972.1 & 2660.8 & 1222.2 & 1023.6 & -11.9 & 311.3 & 15.4 \\
16 & pos.\ $\Gamma$      & WMAP\&SN      & 2971.9 & 2663.9 & 1222.6 & 1024.6 & -10.6 & 308.0 & 13.3 \\
%%%%%%%%%%%%%%%%%%%%%%
%\hline
%WMAP and BAO
%%%%%%%%%%%%%%%%%%%%%%%
17 & best $\Gamma\!=\!0$  & WMAP\&BAO     & 2667.3 & 2660.6 & 1222.2 & 1023.1 & -10.4 & 308.8 & 6.9\\
18 & best $\Gamma\!\neq\! 0$    & WMAP\&BAO     & 2667.7 & 2660.4 & 1221.3 & 1023.7 & -11.5 & 308.1 & 7.3\\
19 & neg.\ $\Gamma$      & WMAP\&BAO     & 2671.7 & 2661.4 & 1221.7 & 1024.2 & -11.7 & 309.3 & 10.3\\
20 & pos.\ $\Gamma$      & WMAP\&BAO     & 2671.7 & 2664.8 & 1224.8 & 1023.9 & -7.7 & 308.3 & 7.0\\
%%%%%%%%%%%%%%%%%%%%%%
%\hline
%WMAP and SN and BAO
%%%%%%%%%%%%%%%%%%%%%%%
21 & best $\Gamma\!=\!0$  & WMAP\&SN\&BAO & 2975.7 & 2660.6 & 1221.8 & 1023.5 & -10.7 & 308.3 & 7.0\\
22 & best $\Gamma\!\neq\! 0$    & WMAP\&SN\&BAO & 2975.8 & 2660.4 & 1220.9 & 1024.0 & -11.7 & 308.3 & 7.1\\
23 & neg.\ $\Gamma$      & WMAP\&SN\&BAO & 2979.3 & 2662.3 & 1220.7 & 1025.6 & -12.1 & 308.7 & 8.2\\
24 & pos.\ $\Gamma$      & WMAP\&SN\&BAO & 2979.6 & 2664.3 & 1222.7 & 1025.5 & -9.9 & 308.0 & 7.3\\
%\hline
%Phenomenology
%%%%%%%%%%%%%%%%%%%%%%%
%\multicolumn{21}{l}{}\\
\multicolumn{1}{l}{} & \multicolumn{9}{l}{Other models appearing in the figures for phenomenological considerations}\\
%\hline
% \multicolumn{4}{|l||}{Model} & $\!\chi^2_{\!\mathrm{WMAP}}$ & $\chi^2_{l<32}$
% & $\chi^{2\mathrm{TT}}$ & $\chi^{2\mathrm{TT}}_{l<32}$ & $\chi^2_{\mathrm{SN}}$ & $\!\chi^2_{\!\mathrm{BAO}}$
% & $\omega_b$ & $\omega_c$ & $H_0$ & $\tau_{re}$ & $\Gamma/H_0$ & $w_e$ &
% $w_0$ & $n_s$ & $A$ & $\Omega_{de0}$ & Age  & $D_{V}^{0.20}$ &  $D_{V}^{0.35}$ & $r_s$ & $z_{\rm dec}$ \\ \hline\hline
%%%%%%%%%%%%%%%%%%%%%%%
25 & \multicolumn{3}{l||}{As Model 1, but $\Gamma/H_0=-0.3$}  & 3680.0 & 1223.6 & 2026.8 & -9.0 & 310.5 & 12.3\\
26 & \multicolumn{3}{l||}{As Model 1, but $\Gamma/H_0=+0.3$}   & 3789.9 & 1224.5 & 2150.0 & -8.8 & 312.7 & 40.3\\
27 & \multicolumn{3}{l||}{As 25, but $\omega_c$ ($\Omega_{de0}$) \& $H_0$ adjusted} & 2661.2 & 1222.1 & 1023.8 & -11.4 & 323.3 & 41.4\\
28 & \multicolumn{3}{l||}{As 26, but $\omega_c$ ($\Omega_{de0}$) \& $H_0$ adjusted} & 2671.4 & 1233.0 & 1022.9 & 0.2 & 307.8 & 10.0\\
\hline
\end{tabular}
%\normalsize
 
%%%%%%%%%%%%%%%%%%%%%%%%%%%%%%%%%%%%%%%%%%%%%%%%%%%%%%%%%%%%%%%%%%%
%%%%%%%%%%%%%%%%%%%%%%%%%%%%%%%%%%%%%%%%%%%%%%%%%%%%%%%%%%%%%%%%%%%
\vspace{10mm}
\contcaption{The cosmological parameters of the best-fitting models. Today's Hubble parameter $H_0$ is in units km$\,$s$^{-1}\,$Mpc$^{-1}$, the age of the universe is given in Giga years, the distance measure $D_V$ at redshifts $z=0.20$ and $z=0.35$ as well as the sound horizon at decoupling, $r_s$, are in units of
$h^{-1}\,$Mpc, where $h$ is defined by $H_0 = h\,$km$\,$s$^{-1}\,$Mpc$^{-1}$. Ampl. denotes the primordial perturbation amplitude, indeed $\ln(10^{10}A_S^2)$.}
\begin{tabular}{|l|r|r|r|r|r|r|r|r|r||r|r|r|r|r|r|}\hline
Model & $\omega_b$
& $\omega_c$ & $H_0$ & $\tau_{re}$ & $\Gamma/H_0$ & $w_e$ & $w_0$ &
$n_s$ & Ampl. & $\Omega_{de0}$ & Age & $D_{V}^{0.20}$ &
$D_{V}^{0.35}$ & $r_s$ & $z_{\rm dec}$ \\ \hline
%\hline
%WMAP
%%%%%%%%%%%%%%%%%%%%%%%
 1 & 0.0229 & 0.107 & 61.1 & 0.09 & 0 & -0.16 & -0.98 & 0.975 & 3.06 & 0.65 & 14.2 & 552.3 & 902.9 & 94.2 & 1017 \\
 2 & 0.0224 & 0.133 & 49.7 & 0.08 & -0.30 & -0.62 & -0.23 & 0.954 & 3.03 & 0.37 & 14.6 & 507.9 & 799.2 & 77.2 &  1017 \\
 3 & 0.0229 & 0.196 & 47.1 & 0.10 & -0.90 & -0.66 & -0.19 & 0.971 & 3.06 & 0.01 & 14.7 & 500.6 & 780.9 & 73.2 &  1017 \\
 4 & 0.0233 & 0.091 & 58.7 & 0.09 & 0.12 & -0.13 & -0.81 & 0.996 & 3.04 & 0.67 & 14.4  & 544.4 & 884.4 & 91.2 &  1018 \\
%%%%%%%%%%%%%%%%%%%%%%
%\hline
%SN
%%%%%%%%%%%%%%%%%%%%%%%
 5 & 0.0757 & 0.013 & 68.8 & 0.08 & 0 & 0.30 & -0.98 & 0.970 & 3.04 & 0.81 & 13.3  & 566.2 & 937.3 & 13.0 &  1130\\
 6 & 0.0338 & 0.083 & 74.6 & 0.08 & 1.85 & -0.84 & -0.96 & 0.970 & 3.04 & 0.79 & 11.3 & 565.9 & 938.0 & 75.5 & 1050 \\
%%%%%%%%%%%%%%%%%%%%%%
%\hline
%BAO
%%%%%%%%%%%%%%%%%%%%%%%
 9 & 0.0457 & 0.010 & 68.3 & 0.08 & 0 & -0.94 & -0.99 & 0.970 & 3.04 & 0.88 & 17.4 & 582.3 & 992.6 & 111.7 & 1050 \\
10 & 0.0079 & 0.010 & 89.7 & 0.08 & 2.92 & -0.87 & -0.98 & 0.970 & 3.04 & 0.98 & 12.8 & 592.7 & 1025 & 116.0 & 989 \\
%%%%%%%%%%%%%%%%%%%%%%
%\hline
%WMAP and SN
%%%%%%%%%%%%%%%%%%%%%%%
13 & 0.0225 & 0.109 & 67.5 & 0.08 & 0 & -0.72 & -0.95 & 0.960 & 3.05 & 0.71 & 13.9  & 559.2 & 924.7 & 104.0 & 1017 \\
14 & 0.0227 & 0.116 & 67.5 & 0.08 & -0.08 & -0.72 & -1.00 & 0.963 & 3.04 & 0.70 & 13.9 & 560.4 & 927.3 & 104.3 &  1017 \\
15 & 0.0223 & 0.133 & 62.4 & 0.09 & -0.25 & -0.61 & -0.99 & 0.955 & 3.05 & 0.60 & 14.1 & 550.2 & 900.0 & 96.8 &  1017 \\
16 & 0.0236 & 0.084 & 64.6 & 0.11 & 0.19 & -0.06 & -1.00 & 1.005 & 3.06 & 0.74 & 14.1 & 561.6 & 926.3 & 99.7 & 1019 \\
%%%%%%%%%%%%%%%%%%%%%%
%\hline
%WMAP and BAO
%%%%%%%%%%%%%%%%%%%%%%%
17 & 0.0226 & 0.107 & 70.2 & 0.09 & 0 & -0.78 & -0.99 & 0.959 & 3.04 & 0.74 & 13.8   & 562.6 & 933.3 & 106.4 &  1017 \\
18 & 0.0226 & 0.111 & 68.1 & 0.08 & -0.04 & -0.66 & -0.99 & 0.963 & 3.04 & 0.71 & 13.9 & 561.5 & 930.0 & 105.4 & 1017 \\
19 & 0.0220 & 0.125 & 64.1 & 0.09 & -0.21 & -0.70 & -0.97 & 0.944 & 3.01 & 0.64 & 14.1  & 553.6 & 909.8 & 100.7 &  1015 \\
20 & 0.0228 & 0.089 & 68.9 & 0.09 & 0.13 & -0.40 & -0.99 & 0.969 & 3.03 & 0.76 & 13.9  & 565.1 & 938.2 & 107.7 &  1017 \\
%%%%%%%%%%%%%%%%%%%%%%
%\hline
%WMAP and SN and BAO
%%%%%%%%%%%%%%%%%%%%%%%
21 & 0.0227 & 0.107 & 68.8 & 0.09 & 0 & -0.68 & -0.99 & 0.964 & 3.05 & 0.73 & 13.8 & 562.6 & 933.3 & 106.4 &  1017 \\
22 & 0.0229 & 0.113 & 69.1 & 0.09 & -0.07 & -0.74 & -1.00 & 0.964 & 3.05 & 0.72 & 13.8  & 562.6 & 933.4 & 107.1 &  1017 \\
23 & 0.0223 & 0.120 & 65.9 & 0.09 & -0.21 & -0.71 & -0.95 & 0.956 & 3.01 & 0.67 & 14.0 & 555.5 & 915.4 & 103.9 &  1016 \\
24 & 0.0233 & 0.089 & 67.9 & 0.09 & 0.13 & -0.34 & -0.97 & 0.983 & 3.03 & 0.76 & 13.9  & 562.7 & 931.8 & 105.8  & 1018 \\
%\hline
%Phenomenology
%%%%%%%%%%%%%%%%%%%%%%%
%\multicolumn{21}{l}{}\\
\multicolumn{1}{l}{} & \multicolumn{15}{l}{Other models appearing in the figures for phenomenological considerations}\\
%\hline
% \multicolumn{4}{|l||}{Model} & $\!\chi^2_{\!\mathrm{WMAP}}$ & $\chi^2_{l<32}$
% & $\chi^{2\mathrm{TT}}$ & $\chi^{2\mathrm{TT}}_{l<32}$ & $\chi^2_{\mathrm{SN}}$ & $\!\chi^2_{\!\mathrm{BAO}}$
% & $\omega_b$ & $\omega_c$ & $H_0$ & $\tau_{re}$ & $\Gamma/H_0$ & $w_e$ &
% $w_0$ & $n_s$ & $A$ & $\Omega_{de0}$ & Age  & $D_{V}^{0.20}$ &  $D_{V}^{0.35}$ & $r_s$ & $z_{\rm dec}$ \\ \hline\hline
%%%%%%%%%%%%%%%%%%%%%%%
25 &  0.0229 & 0.107 & 61.1 & 0.09 & -0.30 & -0.16 & -0.98 & 0.975 & 3.06 & 0.65 & 14.6  & 552.9 & 905.6 & 99.1 & 1016 \\
26 &  0.0229 & 0.107 & 61.1 & 0.09 & 0.30 & -0.16 & -0.98 & 0.975 & 3.06 & 0.65 & 13.9  & 551.6 & 900.0 & 89.1 &  1020 \\
27 &  0.0229 & 0.137 & 56.1 & 0.09 & -0.30 & -0.16 & -0.98 & 0.975 & 3.06 & 0.49 & 14.4  & 538.3 & 868.5 & 86.6 &  1017 \\
28 &  0.0229 & 0.080 & 66.1 & 0.09 & 0.30 & -0.16 & -0.98 & 0.975 & 3.06 & 0.76 & 14.0 & 563.1 & 931.1 & 102.0 &   1017 \\
\hline
\end{tabular}
\end{table*}

\subsection{Positive $\Gamma$ and zero-crossing of $\rho_{de}$ \label{app:zerocrossing}}

There is one additional complication in studying the interacting
model and interpreting the marginalized posterior likelihoods. As
pointed out in \citet{Valiviita:2008iv}, in the case of constant
$w_{de}$, any \emph{positive interaction} $\Gamma/H_0 > 0$ would
lead to a zero crossing of $\rho_{de}$, or in other words
$\Omega_{de}$. This means that starting the background calculation
from today's \emph{positive} value, say $\Omega_{de0} \sim 0.7$,
and integrating backward in time, at some moment
$\tau_{de, \mathrm{zc}}$ in the the past $\Omega_{de}(\tau)$ crosses zero and becomes
negative for $\tau < \tau_{de, \mathrm{zc}}$. While we lack deep
understanding of the nature of dark energy we might even accept
this possibility. However, the perturbation equations
(11) and (12) in the companion paper \citep*{CompanionPert},
%(\ref{eq.delta'de_ourQ}) and (\ref{eq.theta'de_ourQ})
have the dark energy density in the denominator,
and therefore become singular at
the moment $\tau_{de, \mathrm{zc}}$. For this reason, as discussed in
\citet{Valiviita:2008iv}, all positive interactions with our type
of interaction are ruled out if $w_{de}$ is constant. However, the
situation changes dramatically when we allow for time varying
$w_{de}$. In viable models $w_{de}$ today is close to $-1$ and
then, as $-4/5 \le w_e < 1/3$, going towards the past makes
$w_{de}$ less negative or even positive. If $w_{de}$ becomes
enough less negative \emph{before} the moment $\tau_{de, \mathrm{zc}}$,
then the zero crossing can actually be avoided. It turns out that
the threshold value, $w_{e, \mathrm{th}}$, depends mildly on all the
background density parameters and strongly on the interaction
$\Gamma/H_0$ and, of course, $w_0$. The larger positive
interaction we have, the larger $w_e$ we need in order to avoid
the zero-crossing of $\rho_{de}$, and hence the singularity of
the perturbation equations. This means that for a given positive
$\Gamma$ (and the background parameters) all the models with $w_e
<  w_{e,\mathrm{th}}(\Gamma/H_0,w_0)$ will be missing from our
Markov chains. There is no similar ``top-hat cut-off'' of models
for negative $\Gamma$. This represents a difficulty in
interpreting the marginalized posterior likelihoods. Let us assume
a completely symmetric situation with respect to $\Gamma=0$. Then
without the cut-off, we would find $50\%$ of the area under our 1d
marginalized posterior for $\Gamma/H_0$ to lie in the negative
$\Gamma$ region, and $50\%$ in the positive $\Gamma$ region.
However, with the cut-off unavoidably in
operation, even if both positive and negative $\Gamma$ models with
$w_e >  w_{e, \mathrm{th}}(|\Gamma|/H_0,w_0)$ led to exactly the
same theoretical predictions (and hence to the same likelihoods), on
the positive $\Gamma$ side the marginalization integral over $w_e$
collects only the volume $1/3 -  w_{e, \mathrm{th}}(\Gamma/H_0,w_0)
\ll 1.133$, while on the negative $\Gamma$ side the volume factor
is $1/3 - (-4/5) = 1.133$. The volume factor in the positive $\Gamma$
side becomes smaller the larger $\Gamma$ is. Therefore, even in
this hypothetical ``symmetric'' situation, the marginalized
likelihood for $\Gamma$ would show a strong ``preference'' for a
negative interaction.

We demonstrate the cut-off effect in Fig.~\ref{fig:VolumeEffect},
which shows samples from our Markov chains from the run with
WMAP\&BAO data. For example, if $\Gamma/H_0 = -0.1$, then the
good-fit region is $-0.75 \lesssim w_e \lesssim -0.2$, whereas for
$\Gamma/H_0 = +0.1$, all the models $-0.8 \lesssim w_e \lesssim
-0.45$ are forbidden because of the zero crossing of $\rho_{de}$.
Fig.~\ref{fig:VolumeEffect} shows also that when $w_e$ is very
negative, we can to some extent compensate this with less negative
$w_0$, as mentioned above.

With negative values of $\Gamma$ we see in Fig.~\ref{fig:VolumeEffect}
another, milder, cut-off of models between
$-4/5 < w_e < -2/3$. Asymptotically with large interactions
this cut-off line approaches $w_e = -2/3$.
From Table~\ref{table:summary} on page \pageref{table:summary}
we can find an explanation for this behaviour.
Although perturbations in the radiation era behave well and we can set
adiabatic initial conditions, there is a rapidly growing non-adiabatic
mode in the matter era, if $-4/5 < w_e < -2/3$. This mode kicks in
faster the stronger the interaction is. In addition, this mode
grows the faster the further away from $w_e = -2/3$ we are.
Therefore, some models with small interaction rate and/or
$w_e$ close enough to $-2/3$ survive.

\section{Best-fitting models}
\label{app:table}

We have collected in Table~\ref{table:Models} $\chi^2$s and parameters of the
best-fitting models from our MCMC runs with various data sets (Models 1--24), as well as the 
example models discussed in Sec.~\ref{sec:phenomenology}
and Figs.~\ref{fig:OtherModels}, \ref{fig:ISWsource}, and \ref{fig:MatterPower} (Models 25--28). It should be noted that Models 6 and 10 (with $w_e < -0.8$)
would be excluded by the CMB data due to the blow-up of perturbations, but since the SN and BAO data probe background quantities only, these models fit them well. As the $\chi^2$ is rather insensitive to $w_e$ almost equally well-fitting models with $w_e > -0.8$ do exist in our chains.
\clearpage

\bibliography{aamnem,InteractingDElikelihoods}
\bibliographystyle{mn2e}
%\bibliography{InteractingDElikelihoods}

\label{lastpage}

\end{document}